\documentclass[12pt,a4paper]{article}

\usepackage{cite}
\usepackage{amsmath}
\def\gfxon{\usepackage[final]{graphicx}}

\gfxon

\begin{document}

\makeatletter
\@addtoreset{equation}{section}
\makeatother
\newcommand{\sect}{\section}
\renewcommand{\theequation}{\thesection.\arabic{equation}}

\makeatletter
\let\old@makecaption=\@makecaption
\def\@makecaption{\small\old@makecaption}
\makeatother

\newcommand{\hypref}[2]{\ifx\href\asklfhas #2\else\href{#1}{#2}\fi}
\newcommand{\remark}[1]{\textbf{#1}}
\newcommand{\figref}[1]{fig. \ref{#1}}
\newcommand{\Figref}[1]{Fig. \ref{#1}}
\newcommand{\tabref}[1]{tab. \ref{#1}}
\newcommand{\indup}[1]{_{\mathrm{#1}}}
\newcommand{\tsum}{{\textstyle\sum}}
\newcommand{\tprod}{{\textstyle\prod}}

\newcommand{\sfrac}[2]{{\textstyle\frac{#1}{#2}}}
\newcommand{\half}{\sfrac{1}{2}}
\newcommand{\quarter}{\sfrac{1}{4}}

\newcommand{\order}[1]{\mathcal{O}(#1)}
\newcommand{\eps}{\varepsilon}
\newcommand{\Lagr}{\mathcal{L}}
\newcommand{\superN}{\mathcal{N}}
\newcommand{\gym}{g_{\scriptscriptstyle\mathrm{YM}}}
\newcommand{\gtwo}{g_2}
\newcommand{\Tr}{\mathop{\mathrm{Tr}}}
\renewcommand{\Re}{\mathop{\mathrm{Re}}}
\renewcommand{\Im}{\mathop{\mathrm{Im}}}
\newcommand{\Li}{\mathop{\mathrm{Li}}\nolimits}
\newcommand{\cdott}{\mathord{\cdot}}
\newcommand{\singlet}{{\mathbf{1}}}

\newcommand{\lrbrk}[1]{\left(#1\right)}
\newcommand{\bigbrk}[1]{\bigl(#1\bigr)}
\newcommand{\vev}[1]{\langle#1\rangle}
\newcommand{\normord}[1]{\mathopen{:}#1\mathclose{:}}
\newcommand{\lrvev}[1]{\left\langle#1\right\rangle}
\newcommand{\bigvev}[1]{\bigl\langle#1\bigr\rangle}
\newcommand{\bigcomm}[2]{\big[#1,#2\big]}
\newcommand{\lrabs}[1]{\left|#1\right|}
\newcommand{\abs}[1]{|#1|}

\newcommand{\nn}{\nonumber}
\newcommand{\nln}{\nonumber\\}
\newcommand{\nl}{\nonumber\\&&\mathord{}}
\newcommand{\nle}{\nonumber\\&=&\mathrel{}}
\newcommand{\eq}{\mathrel{}&=&\mathrel{}}
\newenvironment{myeqnarray}{\arraycolsep0pt\begin{eqnarray}}{\end{eqnarray}\ignorespacesafterend}
\newenvironment{myeqnarray*}{\arraycolsep0pt\begin{eqnarray*}}{\end{eqnarray*}\ignorespacesafterend}

\newcommand{\NPB}[3]{{\it Nucl.\ Phys.} {\bf B#1} (#2) #3}
\newcommand{\CMP}[3]{{\it Commun.\ Math.\ Phys.} {\bf #1} (#2) #3}
\newcommand{\PRD}[3]{{\it Phys.\ Rev.} {\bf D#1} (#2) #3}
\newcommand{\PLB}[3]{{\it Phys.\ Lett.} {\bf B#1} (#2) #3}
\newcommand{\JHEP}[3]{{JHEP} {\bf #1} (#2) #3}
\newcommand{\hepth}[1]{{\tt hep-th/#1}}
\newcommand{\ft}[2]{{\textstyle\frac{#1}{#2}}}
\newcommand{\cO}{{\cal O}}
\newcommand{\cT}{{\cal T}}
\def\ss{\scriptstyle}
\def\st{\scriptstyle}
\def\sst{\scriptscriptstyle}
\def\ra{\rightarrow}
\def\lra{\longrightarrow}
\newcommand\Zb{\bar Z}
\newcommand\bZ{\bar Z}
\newcommand\bF{\bar \Phi}
\newcommand\bP{\bar \Psi}

\let\oldPhi=\Phi
\let\oldGamma=\Gamma
\renewcommand{\Phi}{\mathnormal{\oldPhi}}
\renewcommand{\Gamma}{\mathnormal{\oldGamma}}

\thispagestyle{empty}
\begin{flushright}
{\sc\footnotesize hep-th/0208178}\\
{\sc\footnotesize AEI 2002-061}
\end{flushright}
\setcounter{footnote}{0}
\begin{center}
{\Large{\bf BMN Correlators and Operator Mixing in 
\boldmath$\superN=4$ Super Yang-Mills Theory}\par
}\vspace{14mm}
{\sc N. Beisert$^a$, C. Kristjansen$^{b,}$\footnote{
Work supported by the Danish Natural Science Research Council.},  
J. Plefka$^a$, G. W. Semenoff$^{c,}$\footnote{
Work supported 
in part by NSERC of Canada.} and M. Staudacher$^a$}\\[4mm]
{\it $^a$Max-Planck-Institut f\"ur Gravitationsphysik\\
Albert-Einstein-Institut\\
Am M\"uhlenberg 1, D-14476 Golm, Germany}\\ [2mm]
{\it $^b$The Niels Bohr Institute\\
Blegdamsvej 17, Copenhagen \O, DK2100 Denmark}\\[2mm]
{\it $^{c}$Department of Physics and Astronomy\\
University of British Columbia\\
Vancouver, British Columbia V6T 1Z1, Canada}\\[2mm]
{\tt nbeisert@aei.mpg.de, kristjan@alf.nbi.dk, plefka@aei.mpg.de,
semenoff@physics.ubc.ca, matthias@aei.mpg.de}\\[2mm]

{\sc Abstract}\\[2mm]
\end{center}

Correlation functions in perturbative $\superN=4$ supersymmetric 
Yang-Mills theory are examined in the Berenstein-Maldacena-Nastase (BMN) limit.
We demonstrate that non-extremal four-point functions of chiral primary
fields are ill-defined in that limit. This lends support to the assertion
that only gauge theoretic two-point functions should be compared to
pp-wave strings.  We further refine the analysis of the recently
discovered non-planar corrections to the planar BMN limit. In particular,
a full resolution to the genus one operator mixing problem is
presented, leading to modifications in the map between BMN operators and 
string states. We give a perturbative
construction of the correct operators and we identify their anomalous
dimensions.  We also distinguish symmetric, anti-symmetric and singlet
operators and find, interestingly, the same torus anomalous dimension for
all three. Finally, it is discussed how operator mixing effects modify
three point functions at the classical level and, at one loop, allow
us to recover conformal invariance.

\newpage
\setcounter{page}{1}
\sect{Introduction and overview}
\label{sec:intro}

Recently, a very interesting proposal for taking a novel kind of
large $N$ limit in a gauge theory was made \cite{Berenstein:2002jq}.
The proposal is to consider correlation functions of 
gauge invariant operators 
with a large SO(2) charge $J$ in $\superN=4$ Super Yang-Mills theory,
where SO(2) is a subgroup of the full SO(6) R-symmetry group of this
gauge theory. This Berenstein-Maldacena-Nastase (BMN) limit is then
\begin{equation}
N \to \infty ~~{\rm and}~~ J\to\infty~~{\rm with} ~~
\frac{J^2}{N}~~{\rm fixed},~~\gym~~{\rm fixed}
\label{limit}
\end{equation}
where $N$ is the rank of the U$(N)$ gauge group. It appears, but has not
been rigorously proven, that the limit is insensitive to the difference
between SU$(N)$ and U$(N)$. The limit is interesting in its own right,
as a large $N$ limit different from the usual `t~Hooft limit. 
The difference is that the latter takes $\gym$ to
zero while holding $\lambda=\gym^2 N$ fixed. In the BMN
limit we are instructed to \emph{not} take $\gym$ to
zero. Let us stress that therefore
the BMN limit is also a priori inequivalent to taking the 
strong coupling limit $\lambda \rightarrow \infty$ 
of the `t Hooft limit together with $J \sim \sqrt{\lambda}$.
Naively, the BMN limit would not be expected to be meaningful in the
quantum gauge theory since every quantum correction involves an
extra factor of $\lambda=\gym^2 N$, which diverges in
the limit \eqref{limit}. This objection trivially does not apply to
protected operators. These are given, in the scalar field sector,
by SO(6) symmetric and traceless combinations of the scalar fields.
A crucial insight led the authors of \cite{Berenstein:2002jq} to
consider operators which violate this symmetry in a small, controlled
fashion. For these BMN operators, with large SO(2) charge $J$,
 quantum corrections were argued to instead be proportional to
\begin{equation}
\lambda'=\frac{\lambda}{J^2}=\frac{\gym^2 N}{J^2}
\label{lambdaprime}
\end{equation}
which is finite in the limit \eqref{limit}. This was first shown
in \cite{Berenstein:2002jq} for the one-loop two-point function of BMN
operators; the analysis was later extended to two loops 
\cite{Gross:2002su}, supporting arguments to all orders in 
perturbation theory were presented in 
\cite{Berenstein:2002jq,Gross:2002su}, and
a proof was proposed in \cite{Santambrogio:2002sb}.
The planar one-loop correction to certain three-point functions of BMN
operators was obtained in \cite{Chu:2002pd}, and again shown to be
proportional to $\lambda'$ instead of $\lambda$.  

Apart from its intrinsic interest as a non-`t~Hooftian large $N$
limit the excitement created by the work of
\cite{Berenstein:2002jq} is mainly due to the proposal that the
correlators of BMN operators are related, via duality, to
type IIB superstrings quantized on a pp-wave space-time background.
The hope is that one can go far beyond the usual AdS/CFT
correspondence: Firstly, the pp-wave background looks simpler than the
space AdS$_5\times$S$^5$ and is actually obtained from the latter
by taking a limit \cite{Blau:2002dy}. In particular string
quantization becomes feasible \cite{Metsaev:2001bj,Metsaev:2002re}.
Secondly, the BMN prescription relates perturbative gauge theory
results to the spectrum of massive states on the string side.

It is natural to go beyond just comparing the spectrum and to try
to relate string scattering amplitudes to gauge theory correlators.
An apparent puzzle is that the BMN limit \eqref{limit} takes 
$N$ to infinity, which, following 't~Hooft, appears to suppress 
non-planar diagrams. One might thus wonder how to extract 
string interactions, which should be proportional to $1/N$.
The resolution of this puzzle is intriguing:
The limit \eqref{limit} is such that the $1/N^2$ suppression of
non-planar contributions is precisely balanced by corresponding
factors of $J^4$ \cite{Kristjansen:2002bb,Berenstein:2002sa,Constable:2002hw}.
Therefore an effective parameter $\gtwo$ appears
\begin{equation}
\gtwo=\frac{J^2}{N}
\label{g2}
\end{equation}
such that a genus $h$ amplitude in the gauge theory is weighted
by a factor $\gtwo^{2 h}$. This phenomenon is suggestive. However,
finding the precise ``dictionary'' relating in detail
pp-strings and gauge fields has so far proved to be difficult and
remains controversial in the literature. In \cite{Constable:2002hw}
as well as \cite{Verlinde:2002ig} it has been proposed that
the true string coupling constant corresponds, on the gauge side,
not to \eqref{g2}, but to the parameter
\begin{equation}
\gtwo~\sqrt{\lambda'}=\frac{J}{\sqrt{N}}~\gym
\label{true}
\end{equation}
According to this logic, applying the correct dictionary, factors
of $\gtwo$ should \emph{always} be accompanied with matching factors
of $\sqrt{\lambda'}$ in \emph{all} quantities dual to a string theory
amplitude. Whether such a dictionary can really be built is a highly
non-trivial open question. A second controversial issue 
concerns the following
basic question: How does one extract string interaction amplitudes
from the gauge theory. Here the just cited proposals
\cite{Constable:2002hw} and \cite{Verlinde:2002ig} substantially
differ: Constable et.al.~\cite{Constable:2002hw} 
give a somewhat ad hoc prescription that relates the string-vertex 
to a gauge theory \emph{three-point function}. In turn, Verlinde
\cite{Verlinde:2002ig}
argues that only Yang-Mills \emph{two-point functions} have
a string interpretation, and multi-string amplitudes should be extracted
from the two-point functions of appropriately defined multi-trace
operators. Resolving these conflicting scenarios is difficult 
since, despite recent progress 
\cite{Spradlin:2002ar,Gopakumar:2002dq,Kiem:2002xn,Huang:2002wf,
Lee:2002rm,Spradlin:2002rv,Klebanov:2002mp,Huang:2002yt,
Gursoy:2002yy,Chu:2002eu}, finding 
the string amplitudes from string field theory proceeds slowly.
Using first-quantized string theory techniques appears to be complicated
as well: The light-cone gauge quantization of
\cite{Metsaev:2001bj,Metsaev:2002re} yields beautiful results for the
spectrum, but a vertex operator formalism for computing scattering amplitudes
appears difficult to establish.

Here we will not take a firm stand on any of these controversial
issues. Instead, we will push ahead the analysis of correlation
functions of BMN operators. A thorough understanding
on the gauge theory side will surely become part of the left-hand
side of the pages of the sought ``dictionary''.

The present paper is organized as follows: We first study
four-point functions at the classical and one-loop level (chapter \ref{sec:4pt}). Then we
consider genus one, classical and one loop two-point functions (chapter \ref{sec:2pt}).
We end by briefly discussing planar three-point functions (chapter \ref{sec:3pt}). 
This logic is
dictated by the following findings: We study planar four-point functions of 
protected BMN operators and find curious discontinuities already at
the classical level. Considering then the one-loop corrections to
the four-point functions (such corrections are expected even for
protected operators) we find them to be \emph{divergent} in the BMN
limit. This does not exclude that an interesting interpretation for
four-point functions, possibly involving a refinement of the 
original BMN procedure, will eventually be found. However, in our opinion
our result strengthens the proposal of Verlinde \cite{Verlinde:2002ig}
that string interactions should be extracted from two-point
functions as opposed to multi-point functions (see discussion above).
This motivates us to take a fresh look at the torus-level two-point 
functions of BMN operators with impurities. It turns out that the existing
treatment \cite{Kristjansen:2002bb,Constable:2002hw} is incomplete.
In fact\footnote{
Here we would like to acknowledge helpful discussions with M.~Bianchi
and, separately, G. Arutyunov in June 2002 on the need to
include double-trace operators for the correct computation of the
torus correction to the anomalous dimension.
A preliminary discussion of operator mixing was also presented in
July by S.Minwalla at Strings 2002.
},
one needs to take into account operator mixing
between single-trace and double-trace operators, as was first proposed
in the literature by Bianchi et.al.~in \cite{Bianchi:2002rw} (page 19).
After implementation of these effects we are able to derive
the correct anomalous dimension $\Delta_n$
of the (redefined) BMN operators. 
The final result (up to ${\cal O}(1/N^2)$)
for the $n$-th two-impurity BMN operator of charge J reads
\begin{equation}
\Delta_n=J+2+
\frac{\lambda'}{8\pi^2}\lrbrk{8\pi^2 n^2+\gtwo^2 
\lrbrk{\frac{1}{6}+\frac{35}{16\pi^2n^2}}
},
\label{upshot}
\end{equation}
where $n \neq 0$. In addition, we extend the definition of BMN operators with
two defects by distinguishing symmetric, anti-symmetric and singlet
operators. The \emph{same} torus anomalous dimension \eqref{upshot} is found
for all three types of operators. 
We should mention in passing that the expression eq.\eqref{upshot} 
does not quantitatively agree with the result 
obtained in \cite{Verlinde:2002ig}. 

The fact that the original, 
single trace BMN operators have to be modified by double trace
operators also has an interesting influence on three-point functions,
as will be discussed in the final chapter. Three-point functions
of impurity BMN operators were first considered classically in
\cite{Constable:2002hw} and, for special cases, 
at one-loop in \cite{Chu:2002pd}.
Actually, for the general one-loop three-point function of the
original BMN operators one finds\footnote{
\label{fn:Chu}
C-S~Chu kindly informed us about this (unpublished) result.}
that the result is inconsistent with conformal invariance. 
As we shall show this
problem is resolved by 
using the redefined operators%
\footnote{
In fact, 
the requirement of a consistent one-loop three-point function
is another way to obtain the correct operator redefinition.}.
Interestingly, this obligatory redefinition of operators not only 
changes the result at one-loop, but also at the classical level.
The planar, free three-point functions of
\cite{Constable:2002hw}, while being quite important as an auxiliary tool
for finding the correct operator mixing, are therefore
seen to lack physical significance. As a consequence, the
reported agreement of these ``bare'' three-point functions
with string field theory calculations
\cite{Constable:2002hw,Huang:2002wf,
Spradlin:2002rv} seems to indicate that the latter
have to be reconsidered as well.

\sect{Four-point functions in the BMN limit}
\label{sec:4pt}

\subsection{General remarks}
\label{ssec:4ptintro}

So far the existence of the BMN limit has only been tested in
the literature for various two- and three-point functions of the operators
proposed by \cite{Berenstein:2002jq}. However, in
\cite{Verlinde:2002ig} it was argued, from a slightly different point
of view, that only gauge theory two-point functions possess an
interpretation on the string theory side. Indeed, comparing the
string quantization in a pp-wave background to the gauge theory limit,
one immediately faces a puzzle, even before embarking on any concrete
calculations of multi-point functions: pp-strings in light-cone gauge
are confined in the eight transverse directions by harmonic oscillator
potentials and, therefore, propagating transverse zero-modes do not
exist. However, if we were allowed to place BMN operators on
arbitrary space-time points we would expect such translationally
invariant zero-modes. Space-time seems to have disappeared\footnote{
For a related, recent discussion in a simpler setting see
\cite{Arutyunov:2002xd}.} in the BMN
limit! This problem can be (and has been) ignored for two- and
three-point functions: The functional form of such correlators is
fixed by conformal invariance, and involves only powers of the 
distances between points. For two-point functions, these powers are 
used to extract scaling dimensions (which are then related 
to the energies of the corresponding string states) but the
space-time factor is ignored otherwise. For three-point
functions, a similar, heuristic \cite{Constable:2002hw} procedure
relates the string interaction vertex to the numerical coefficient
of the Yang-Mills correlation function, 
while the space time factors are simply dropped.
It is well-known that conformal invariance no longer fixes the
space-time form of four- and higher-point correlation functions:
These depend, in a priori complicated ways, on conformal
ratios of the space-time differences. In \cite{Chu:2002qj} it has been
suggested that gauge theory correlators should nevertheless be related to
$n$-string amplitudes by taking short-distance (pinching) limits
of the gauge theory $n$-point functions. However,
this procedure appears to assume, firstly, certain analytic 
properties of the amplitudes, such as \emph{continuity} 
as one brings space-time
points together, and, secondly, the \emph{existence} of the BMN limit
of the multi-point function for separated, if close, points.

We will now study these issues in the simplest non-trivial setting:
The (non-extremal), connected correlation function of four chiral
primary operators. In the course of the analysis, we shall,
interestingly, find that both of the above assumptions (continuity and
existence) fail in this setting.
The operators to be studied are \cite{Berenstein:2002jq}
\begin{equation}
\cO^J(x) = \frac{1}{\sqrt{J}}\,\lrbrk{\frac{8\pi^2}{\gym^2 N}}^{J/2}\, \Tr \, Z^J(x),
\label{defchiral}
\end{equation}
where $Z=\frac{1}{\sqrt{2}}(\phi_5+i \phi_6)$ is the complex 
superposition of two of the six real scalar fields of the model. 
For details of our notation see appendix \ref{ssec:notation}.
They are, to leading, i.e.~planar order, conjectured to correspond to
the ground states $|0,p^+ \rangle$ of the light-cone pp-string.
We shall consider the connected four-point function
\begin{equation}
G^{J_1 J_2,K_1 K_2}_{x_1 x_2,y_1 y_2}=\bigvev{\cO^{J_1}(x_1)\cO^{J_2}(x_2)
\bar \cO^{K_1}(y_1) \bar \cO^{K_2}(y_2)}\indup{conn}
\label{fourpoint}
\end{equation}
with $J_1+J_2=J=K=K_1+K_2$.
One may, without loss of generality, assume $0 < J_1\leq K_1,K_2\leq
J_2$. We will begin by studying this correlator in the planar limit, first
classically and then including one-loop radiative corrections.
Finally we will present and discuss the double-scaled free field
theory result for eq.\eqref{fourpoint}.

\subsection{Planar, free field theory result}
\label{ssec:4ptplanarfree}

\begin{figure}
\centering
\parbox[b]{3.5cm}{\centering\includegraphics{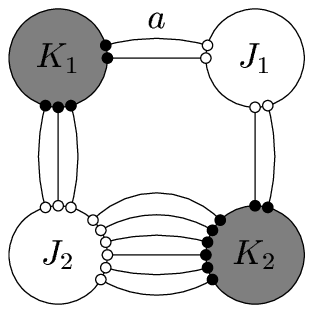}}\quad
\parbox[b]{4.5cm}{\centering\includegraphics{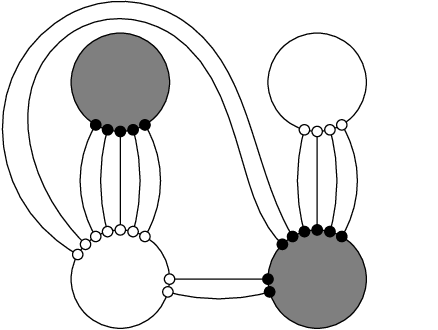}}\quad
\parbox[b]{3.5cm}{\centering\includegraphics{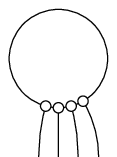}}\\
\parbox{3.5cm}{\centering 1 perm.}\quad
\parbox{4.5cm}{\centering 4 perm.}\quad
\parbox{3.5cm}{\centering 4 perm.}

\caption{Spherical diagrams for the four-point correlator.
The charges of the depicted operators 
are $J_1=4$, $J_2=9$, $K_1=5$, $K_2=8$.
In the left diagram the number of lines, $a=2$, 
between $J_1$ and $K_1$ is not fixed.
In the middle diagram the distribution of lines between
$J_2$ and $K_2$ is not fixed. The right diagram
is unique.
Furthermore, permutations of the four operators must be taken into account.
}
\label{fig:PP.22Sphere}
\end{figure}

Unlike the case of two- or three-point functions, the space-time
dependence of eq.\eqref{fourpoint} is no longer fixed by 
conformal invariance. This can already be seen in free field theory.
Consider \figref{fig:PP.22Sphere}, which illustrates the three distinct types of
possible planar Wick contractions of the four operators. It is
clear that we have to sum over the number $a$ of contractions connecting
the operators $\cO^{J_1}(x_1)$ and $\bar \cO^{K_1}(y_1)$.
The corresponding space-time weight factor is
\begin{equation}
D^{J_1 J_2,K_1 K_2;a}_{x_1 x_2,y_1 y_2} =
\frac{1}{(x_1-y_1)^{2a} (x_1-y_2)^{2J_1-2a} (x_2-y_1)^{2K_1-2a}
(x_2-y_2)^{2J_2-2K_1+2a}} 
\label{defD}
\end{equation}
which can be also be written as 
\begin{equation}
D^{J_1 J_2,K_1 K_2;a}_{x_1 x_2,y_1 y_2}
=D^{J_1 J_2,K_1 K_2;0}_{x_1 x_2,y_1 y_2}~q^a
\end{equation}
where
\begin{equation}
q=q_{x_1 x_2,y_1 y_2}=\frac{(x_1-y_2)^2(x_2-y_1)^2}{(x_1-y_1)^2(x_2-y_2)^2}.
\label{ratio}
\end{equation}

For the first class of diagrams in \figref{fig:PP.22Sphere} there is one way to 
distribute the lines for a given $a$, so the contribution is 
\begin{equation}
\sum_{a=1}^{J_1-1} D^{J_1 J_2,K_1 K_2;a}_{x_1 x_2,y_1 y_2}
=D^{J_1 J_2,K_1 K_2;0}_{x_1 x_2,y_1 y_2}
\sum_{a=1}^{J_1-1} q^a=D^{J_1 J_2,K_1 K_2;0}_{x_1 x_2,y_1 y_2}\,
\frac{q^{J_1}-q}{q-1}
\end{equation}
or $(J_1-1) D^{J_1 J_2,K_1 K_2;0}_{x_1 x_2,y_1 y_2}$ for $q=1$.
For the second class $a$ must be either $0$ or $J_1$ and there are
$K_2-J_1$ or $K_1-J_1$ lines, respectively, to be distributed on
two multi-lines. The contribution is thus (making use of
$J_1+J_2=K_1+K_2$)
\begin{equation}
(J_2-K_1-1)~D^{J_1 J_2,K_1 K_2;0}_{x_1 x_2,y_1 y_2} +
(J_2-K_2-1)~D^{J_1 J_2,K_1 K_2;J_1}_{x_1 x_2,y_1 y_2}.
\end{equation}
The third class contributes a single diagram for $a=0$ and $a=J_1$, 
i.e. $D^{J_1 J_2,K_1 K_2;0}_{x_1 x_2,y_1 y_2}+
D^{J_1 J_2,K_1 K_2;J_1}_{x_1 x_2,y_1 y_2}$.
Furthermore, there are overall factors of $J_1$, $J_2$, $K_1$, $K_2$ to
account for the different ways to connect the lines to the $Z$s inside 
the traces. Putting everything together, we find, for finite $J$, and 
to leading order in $N$ and $\gym^2$,
the correlator $G^{J_1 J_2,K_1 K_2}_{x_1 x_2,y_1 y_2}$ to equal
\begin{equation}
\frac{\sqrt{J_1 J_2 K_1 K_2}}
{N^2}\,
\lrbrk{\frac{q^{J_1}-q}{q-1}+(J_2-K_1)+(J_2-K_2)~q^{J_1}}
D^{J_1 J_2,K_1 K_2;0}_{x_1 x_2,y_1 y_2}.
\label{freefour}
\end{equation}

We can now take the large $J$ limit, keeping the space-time structure
fixed, as originally proposed in \cite{Berenstein:2002jq}. 
Here we must distinguish the
cases $q<1$, $q=1$ and $q>1$ due to the exponential $q^{J_1}$. 
For $q\leq 1$ we take $D^{J_1 J_2,K_1 K_2;0}_{x_1 x_2,y_1 y_2}$ 
as the space-time factor,
for $q\geq 1$ we take $D^{J_1 J_2,K_1 K_2;J_1}_{x_1 x_2,y_1 y_2}$ 
to absorb the divergent terms $q^{J_1}$,
for $q=1$ both factors actually match.
The large $J$ limit is 
\begin{equation}
G^{J_1 J_2,K_1 K_2}_{x_1 x_2,y_1 y_2}
= \frac{\sqrt{J_1 J_2 K_1 K_2}}{N^2} 
\begin{cases}
\hfill (J_2-K_1)~D^{J_1 J_2,K_1 K_2;0}_{x_1 x_2,y_1 y_2} &\mbox{for }q<1,\\
\hfill J_2~D^{J_1 J_2,K_1 K_2;0}_{x_1 x_2,y_1 y_2}&\mbox{for }q=1,\\
\hfill (J_2-K_2)~D^{J_1 J_2,K_1 K_2;J_1}_{x_1 x_2,y_1 y_2}&\mbox{for }q>1.\\
\end{cases} 
\label{freefourBMN}
\end{equation}
Note that the result for the correlator depends on whether the
conformal ratio $q=q_{x_1 x_2,y_1 y_2}$, which is a continuous 
function of the four positions $x_1$,$x_2$,$y_1$,$y_2$,
is smaller, equal or larger than one. Moreover, this dependence
is non-analytic, and actually \emph{discontinuous}. In particular,
this discontinuity is seen if we consider the pinching 
limit $x_1 \rightarrow x_2$, $y_1 \rightarrow y_2$. 
Nevertheless the discontinuity is not only seen
when pinching: E.g.~one has $q=1$ when the four points are located
at the four corners of a perfect tetrahedron. Upon slightly dislocating
any single one of the operators the correlation will jump. 
We did not investigate the free, planar four-point functions of
BMN operators with impurities. 
However, we believe that the above discontinuities will plague these
operators as well.  

\subsection{Planar, one-loop radiative corrections}
\label{ssec:4ptplanar1loop}

\begin{figure}
\centering
\parbox[b]{4.2cm}{\centering\includegraphics{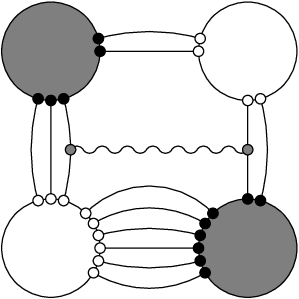}}\quad
\parbox[b]{4.2cm}{\centering\includegraphics{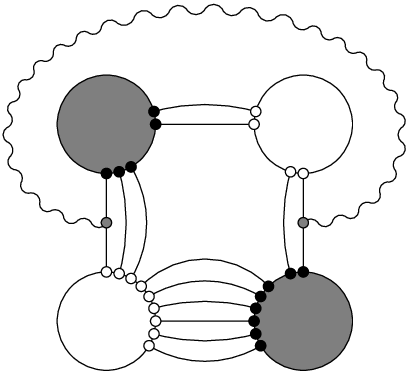}}\vspace{0.5cm}

\parbox[b]{3.5cm}{\centering\includegraphics{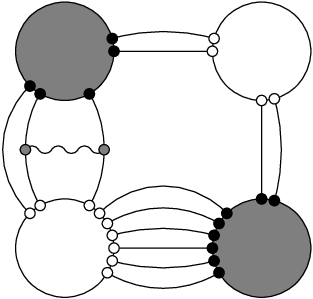}}\quad
\parbox[b]{4.5cm}{\centering\includegraphics{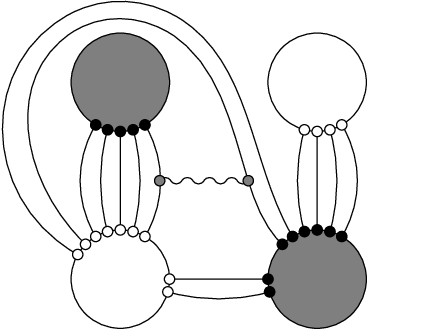}}\quad
\parbox[b]{3.5cm}{\centering\includegraphics{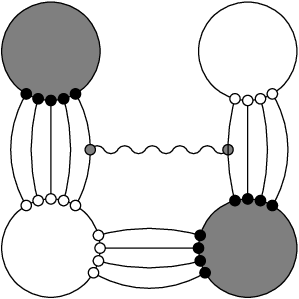}}

\caption{Planar diagrams of four-point correlators
with leading radiative corrections. The wiggly line represents
the sum of \emph{all} radiative contributions within a face of the
diagram, as explained in detail in appendix \ref{app:4pt}.
Diagrams like the ones on the top line contribute
while diagrams on the bottom line do not.
}
\label{fig:PP.22SphereG2}
\end{figure}

Here we will investigate the leading ${\cal O}(\gym^2)$
quantum corrections to the above four-point functions of chiral
primary operators, at the planar level and in the BMN limit.
This is interesting since it is well known that, even though
the operators \eqref{defchiral} are ``protected'', quantum corrections
are only absent at the level of two- and three-point functions.
Four-point functions of such operators are, generically, \emph{not}
protected. The reason is that unprotected operators appear in 
intermediate channels of the correlation functions.
Of course one might have hoped that such corrections are suppressed
in the BMN limit. This will turn out to be not the case. Worse,
we shall find that the quantum corrections are infinite in the BMN
limit.

To compute the planar one-loop correction to the free, planar four-point 
correlator eq.\eqref{freefour} we have to ``decorate'' the 
diagrams of \figref{fig:PP.22Sphere} with either one one-loop self-energy insertion,
one scalar four-point interaction, or one scalar-scalar gluon
exchange.  
Planarity allows to reorganize the diagrammatics such that an effective 
``gluon'' interaction, containing the combined effect of all these
diagrams, affects all scalar lines bounding a face of the free
diagram. 
The details of this calculation are deferred to appendix \ref{app:4pt}.
It is then demonstrated that only the first class of diagrams 
in \figref{fig:PP.22Sphere} has non-vanishing radiative corrections, which result
from the effective interaction of the two quadrangles (four-gons)
of the box-type diagrams. The effective interactions for all
two-gons, as well as those for all faces of the other two classes
of diagrams in \figref{fig:PP.22Sphere}, vanish. 
The situation is schematically illustrated in \figref{fig:PP.22SphereG2}.
%
The result for the quantum correction to the free
correlator eq.\eqref{freefour} (\emph{cf} appendix \ref{app:4pt}) reads
\begin{equation}
\delta G^{J_1 J_2,K_1 K_2}_{x_1 x_2,y_1 y_2}
=-\frac{\sqrt{J_1 J_2 K_1 K_2}}{N^2}
\,\frac{\gym^2N}{8\pi^2}\,
\frac{q^{J_1}-q}{q-1}\,
\Phi(r,s)\,
D^{J_1 J_2,K_1 K_2;0}_{x_1 x_2,y_1 y_2}~
\label{qfour}
\end{equation}
where $D^{J_1 J_2,K_1 K_2;0}_{x_1 x_2,y_1 y_2}$ is given, as before, by
eq.\eqref{defD} (with $a=0$) and the conformal ratio 
$q$ by eq.\eqref{ratio}.
The function $\Phi(r,s)$, whose explicit form is given in appendix \ref{app:4pt}, 
is a complicated but finite function
of the remaining conformal ratios $r,s$ (only two of the three ratios $q,r,s$
are independent):
\begin{equation}
r=r_{x_1 x_2,y_1 y_2}=
\frac{(x_1-y_1)^2 (x_2-y_2)^2}{(x_1-x_2)^2 (y_1-y_2)^2},
\quad
s=s_{x_1 x_2,y_1 y_2}=
\frac{(x_1-y_2)^2 (x_2-y_1)^2}{(x_1-x_2)^2 (y_1-y_2)^2}.
\end{equation}
Now we are ready to investigate the the BMN limit 
$J \rightarrow \infty$ of this leading radiative correction.
As for the free case we find a discontinuity at $q=1$:
%
\begin{equation}
\label{qfourBMN}
\delta G^{J_1 J_2,K_1 K_2}_{x_1 x_2,y_1 y_2}=
-\frac{\sqrt{J_1 J_2 K_1 K_2}}{N^2} 
\,\frac{\gym^2}{8\pi^2}\,
\Phi(r,s)~N
\begin{cases}
\frac{q}{1-q}~D^{J_1 J_2,K_1 K_2;0}_{x_1 x_2,y_1 y_2} &\mbox{for }q<1,\\
\,\,\, J_1~D^{J_1 J_2,K_1 K_2;0}_{x_1 x_2,y_1 y_2}&\mbox{for }q=1,\\
\frac{1}{q-1}~D^{J_1 J_2,K_1 K_2;J_1}_{x_1 x_2,y_1 y_2}&\mbox{for }q>1.\\
\end{cases}  
\end{equation}
Actually, the discontinuity is now even worse since the power of
$J$ changes in a discrete fashion as $q \rightarrow 1$.
Furthermore, we see that the quantum correction
$\delta G^{J_1 J_2,K_1 K_2}_{x_1 x_2,y_1 y_2}$ 
\emph{diverges} relative to the classical contribution 
$G^{J_1 J_2,K_1 K_2}_{x_1 x_2,y_1 y_2}$ 
in the BMN limit: The extra power of $N$ in eq.\eqref{qfourBMN}
scales as $N \sim J^2$.
For $q=1$ the divergence is therefore quadratic in $J$, 
while for $q \neq 1$ it is linear in $J$.

It is interesting to check the ``double-pinching'' limit
$x_1 \rightarrow x_2$, $y_1 \rightarrow y_2$ 
of eq.\eqref{qfour} \emph{before} taking the BMN limit:
In this case the function $\Phi(r,s)$ vanishes,
and there are no quantum corrections at all.
This is easy to understand, since we then effectively
compute a two-point function of protected double-trace operators
$\bigvev{\cO^{J_1}\cO^{J_2}(x_1) \bar \cO^{K_1} \bar \cO^{K_2}(y_1)}$.

One expects the above result to also hold for
BMN operators with impurities. 
We conclude that the BMN limit of $\superN=4$ perturbative
Super Yang-Mills theory does not appear to be meaningful
for general correlation functions of BMN operators. The above
discontinuity and divergence properties cast some doubts on attempts
to relate Yang-Mills $n$-point functions to $n$-string
amplitudes. This is consistent with the picture proposed in 
\cite{Verlinde:2002ig} that one should only compare two-point
functions of multi-trace operators to multi-string amplitudes.
We feel that our result also questions the validity of 
the proposal
of \cite{Constable:2002hw} that relates the Yang-Mills 
three-point function to the string three-vertex: The information
obtained from the three-field correlator of BMN operators
can also be obtained by pinching the three-point function 
before taking the BMN limit
and subsequently extracting the amplitude from the resulting two-point
function. This is again in accordance with the proposal of 
\cite{Verlinde:2002ig}.

\subsection{Double-scaled free field theory result 
(at {\mathversion{bold}$q=1$})}
\label{ssec:4ptmm}

So far we have only considered the correlation function
eq.\eqref{fourpoint} in the strict planar limit, as in
\cite{Berenstein:2002jq}, exposing two types of obstructions to
a meaningful BMN limit for four-point functions. 
For completeness, we would like to discuss the structure of 
\emph{free, non-planar} contributions to these correlators.
As originally shown in \cite{Kristjansen:2002bb}, these are
generically finite and non-vanishing in the BMN limit.
This remains true for the four-point function eq.\eqref{fourpoint}.
Indeed, using the methods of \cite{Kristjansen:2002bb}, 
it is straightforward to work out the non-planar, free-field
corrections to eq.\eqref{freefourBMN}. E.g.~the torus correction reads
\begin{equation}
\delta_{\frac{1}{N^2}} G^{J_1 J_2,K_1 K_2}_{x_1 x_2,y_1 y_2}
=G^{J_1 J_2,K_1 K_2}_{x_1 x_2,y_1 y_2}~\frac{1}{24N^2}
\begin{cases}
\frac{J (K_1^4-K_2^4)+2 J_1 J_2 (K_1^3-K_2^3)}{K_1-K_2} &\mbox{for } q\neq 1\\
J_1^2 J^2+J_2^2 (K_1^2+K_2^2)&\mbox{for }q=1.\\
\end{cases} 
\label{freetorus}
\end{equation}
We observe that the intriguing discontinuities found above continue to
be present at ${\cal O}(\frac{1}{N^2})$.
For $q=1$ we can, extending our results in \cite{Kristjansen:2002bb}, 
go further and find, using matrix model techniques,
the complete $\frac{1}{N}$ expansion of the free-field contribution
to the four-point function. The result (see appendix~\ref{app:4ptmm}) reads
%
\begin{equation}\label{scaling}
G^{J_1 J_2,K_1 K_2}_{x_1 x_2,y_1 y_2} \Big|^{q=1}_{{\rm free}} =
D^{J_1 J_2,K_1 K_2;0}_{x_1 x_2,y_1 y_2}\,
\frac{4 J}{\sqrt{J_1 J_2 K_1 K_2}}\,
\frac{\sinh  \frac{J J_1}{2 N} \,
\sinh  \frac{J_2 K_1}{2 N}  \,
\sinh  \frac{J_2 K_2}{2 N} }{\frac{J^2}{2 N}} .
\end{equation}
The result shows that the ``scaling functions'', discovered
in \cite{Kristjansen:2002bb}, also appear in the description of
the free field limit of non-extremal four-point functions
of chiral primaries
to all orders in the effective coupling constant $\gtwo=\frac{J^2}{N}$.
A puzzling question is whether all-genus expressions such as
eq.\eqref{scaling} have an interpretation on the string side.
According to \cite{Constable:2002hw,Verlinde:2002ig}
this should not be the case: These authors argue that the true
string interactions always come with a factor 
$\sqrt{\lambda'} \frac{J^2}{N}$ where $\lambda'=\gym^2 \frac{N}{J^2}$,
and that the above free non-planar corrections should be
absorbed in the gauge-string dictionary. For chiral operators,
this dictionary seems to be highly non-unique if we allow for general
operator mixing (see section \ref{ssec:2ptlevelzero}); we feel
that the issue should be understood much better.

\sect{Two-point functions, operator mixing and 
one-loop toroidal anomalous dimensions}
\label{sec:2pt}

\subsection{General remarks}
\label{ssec:2ptintro}

In the previous chapter we demonstrated that the quantum corrections
to the generic space-time correlation function of the 
chiral primary operators 
$\cO^J=\frac{1}{\sqrt{J}}\bigbrk{\frac{8\pi^2}{\gym^2 N}}^{J/2} \Tr Z^J$,
$Z=\frac{1}{\sqrt{2}}(\phi_5+i \phi_6)$,
defined in eq.\eqref{defchiral} are
divergent in the BMN limit \eqref{limit}, 
in line with naive expectation. This problem does not appear at the
level of two- and three-point functions of these operators, 
which are protected against quantum corrections. An interesting
slight modification of the operators $\cO^J$ was invented in 
\cite{Berenstein:2002jq}: If one inserts two ``defect fields'',
e.g.~two of the scalar fields $\phi_i$ ($i=1,2,3,4$) not appearing in 
eq.\eqref{defchiral}, at two arbitrary positions inside the trace,
the resulting operators ($i \neq j$)
\begin{equation}
\cO^{J,p}_{ij} = \lrbrk{\frac{8\pi^2}{\gym^2 N}}^{J/2+1} \Tr (\phi_i Z^p \phi_j Z^{J-p}) 
\qquad {\rm with} \qquad p=0,1, \ldots, J
\label{defect}
\end{equation}
are no longer protected; however, for very large $J$ one could expect
the resulting quantum effects to be small. In free field theory,
and in the planar limit, the operators \eqref{defect} are orthonormal.
Computing their planar two-point function at one loop this is no longer the
case, since interactions can exchange the positions of the fields
$\phi_i$ and the adjacent fields $Z$, leading to operator mixing
between the fields $\cO^{J,p}_{ij}$. However, by a linear
transformation the fields can be diagonalized at the planar and
one-loop level, resulting in the BMN
operators\footnote{
There has been some discussion in the literature concerning the 
detailed definition of these operators. In 
\cite{Kristjansen:2002bb,Constable:2002hw} it was shown that the sum 
should start at $p=0$, as opposed to $p=1$ \cite{Berenstein:2002jq}.
Bianchi et.al.~\cite{Bianchi:2002rw}
proposed to replace the phase factor
$e^{2\pi i p\, n /J}$ by $e^{2\pi i p\, n /(J+1)}$ while the
recent work \cite{Parnachev:2002kk} argues for 
$e^{2\pi i (p+1) n /(J+2)}$
in order to consistently reproduce $\frac{1}{J}$ corrections to the
BMN limit. The latter two modifications do not affect the BMN limit.}
\cite{Berenstein:2002jq}:
\begin{equation}
\cO^{J}_{ij,n}(x) = \frac{1}{\sqrt{J}}
\sum_{p=0}^{J} e^{2\pi i p\, n /J}~\cO^{J,p}_{ij}(x).
\label{olddef}
\end{equation}
After ensuring orthogonality of the operators one may extract their
scaling dimension $\Delta_n$ from the two-point function
\begin{equation}
\bigvev{\cO^J_{ij,m}(x)\,\bar \cO^J_{ij,n}(0)}=
\frac{\delta_{mn}}{|x|^{2 \Delta_n}}.
\label{defDelta}
\end{equation}
It is anomalous since for $n  \neq 0$ it deviates in the quantum
theory by $(\delta \Delta_n)$ from the classical dimension $J+2$: 
\begin{equation}
\Delta_n=J+2+(\delta \Delta_n).
\label{defdeltaDelta}
\end{equation}
At one loop, and in the planar limit, one has \cite{Berenstein:2002jq}
\begin{equation}
(\delta \Delta_n)=\lambda' n^2
\label{bmndimension}
\end{equation}
where $\lambda'$ is finite in the BMN limit, 
see \eqref{limit},\eqref{lambdaprime}.
The orthogonalization \eqref{olddef} appears to be valid, at large
$J$, and at the planar level, 
to all orders in the coupling, and the exact\footnote{
It has been pointed out to us by G.Arutyunov that it is
far from obvious that the operators defined in 
eq.\eqref{olddef} are exact eigenstates of the dilatation operator
of the conformal field theory. They certainly aren't at finite $J$;
in particular they mix between various different irreducible
representations of the superconformal group. However, one might
conjecture that these subtleties are irrelevant in the 
``continuum limit'' $J \rightarrow \infty$.}
planar anomalous dimension is believed to be known 
\cite{Berenstein:2002jq,Gross:2002su,Santambrogio:2002sb}.
Technically, the one-loop anomalous dimension is obtained as follows.
Computing in one-loop perturbation theory the correction to
the free result one finds
\begin{equation}
\bigvev{\cO^J_{ij,m}(x)\,\bar \cO^J_{ij,n}(0)}=
\frac{\delta_{mn}}{|x|^{2J+4} } \,
\bigbrk{1+ (\delta \Delta_n)~L}
\label{extract}
\end{equation}
where $L=\log (x\Lambda)^{-2}$.
Clearly $L$ reproduces the expected $x$-dependence
when expanding the definition eq.\eqref{defDelta}, 
using eq.\eqref{defdeltaDelta}; 
$\log \Lambda^2$ is a divergent constant (depending
on the regularization scheme employed) that sets the scale.
More details can be found in appendix \ref{app:effver}.

For the remainder of this chapter we will shorten the notation
by omitting the $x$-dependence of the correlators
as well as the multiplicative factors $\frac{\gym^2}{8 \pi^2}$
from the scalar propagators and operator normalizations 
in \eqref{defect}.
In these conventions eq.\eqref{extract} compactly reads
\begin{equation}
\bigvev{\cO^J_{ij,m} \bar \cO^J_{ij,n} }=
\delta_{mn}~\bigbrk{1+ (\delta \Delta_n)~L}.
\label{extract2}
\end{equation}
 
\begin{figure}
\centering
\includegraphics{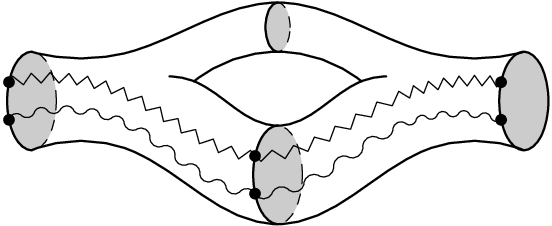}\qquad
\includegraphics{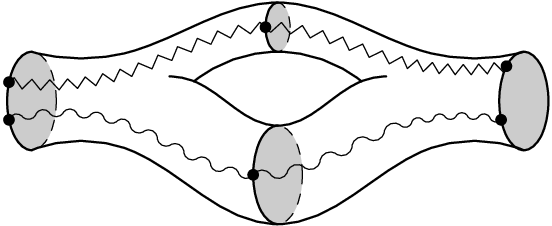}

\caption{Cutting the torus of a correlator
between two single trace operators yields double trace operators.
The two impurities can either 
be on the same branch or on different branches
corresponding to ${\cal T}^{J,r}_{ij,n}$ and ${\cal T}^{J,r}_{ij}$
of \eqref{double},
respectively.}
\label{fig:PP.TorusCut}
\end{figure}

It is natural to ask for the non-planar corrections to the
anomalous dimensions eq.\eqref{bmndimension}. Important
steps in this direction were undertaken in 
\cite{Kristjansen:2002bb,Constable:2002hw},
where the genus one classical and one-loop quantum corrections
to the correlator eq.\eqref{extract2} were computed.
The result reads\footnote{
Our result in \cite{Kristjansen:2002bb}
differs from the analogous expressions eqs.(4.9),(4.10) in
\cite{Constable:2002hw} in one important respect. 
In fact, the last term in each of these equations should come with
the \emph{opposite} sign in order to agree with the correct 
eq.\eqref{startingpoint}. 
In particular, the contributions in question \emph{increase} 
the planar anomalous dimensions
$\lambda' n^2$, in contradistinction to the decrease found in
\cite{Constable:2002hw}. This also means that the unitarity check in 
section 5.2 of \cite{Constable:2002hw} appears to fail due to the differing sign.
},
putting e.g.~$i=1,j=2$
\begin{myeqnarray}
\bigvev{\cO^J_{12,m}\, \bar\cO^J_{12,n}} \eq
\delta_{mn}
\lrbrk{1+\frac{\lambda'L}{8\pi^2}\,8\pi^2 m^2}
\nl
+\gtwo^2\,\lrbrk{M^1_{mn}
+\frac{\lambda' L}{8\pi^2}\lrbrk{8\pi^2mn M^1_{mn}+{\mathcal{D}^1_{mn}}}}
\nl
+\order{\gtwo^4}
\label{startingpoint}
\end{myeqnarray}
where the matrices $M^1_{mn}$, ${\mathcal{D}^1_{mn}}$ can be found
in appendix \ref{app:matel} 
(the notation is the one of \cite{Kristjansen:2002bb},
except for the additional upper index $1$ on the matrices in the
present paper). 
As was already discussed in detail in 
\cite{Kristjansen:2002bb,Constable:2002hw}, on the torus
the operators eq.\eqref{olddef} are no longer orthogonal classically:
The matrix $M^1_{mn}$ is not diagonal. 
By a linear transformation we could proceed to (\textit{i}) orthonormalize the
classical contribution in eq.\eqref{startingpoint} and (\textit{ii})
subsequently diagonalize the quantum contribution
by an orthogonal transformation; this would not affect the 
classical part which would be already proportional to the unit matrix 
$\delta_{mn}$ after
step (\textit{i}). This is nevertheless \emph{not} correct.
The reason is easy to understand pictorially:
Considering the torus correction to a two-point function, we see
from \figref{fig:PP.TorusCut} that double-trace operators appear 
in intermediate channels.
And indeed, as we shall find in the next section, 
the overlap between such double-trace operators and the single-trace 
BMN operators is of ${\cal O}(\gtwo)$. It therefore affects 
the ${\cal O}(\gtwo^2)$ anomalous dimension upon diagonalization.
We conclude that the calculations of 
\cite{Kristjansen:2002bb,Constable:2002hw}
are not quite complete, and we will now proceed to
derive the correct dimensions.

\subsection{Operator definitions}
\label{ssec:2ptopdef}

Let us recapitulate and slightly extend the definitions of the 
properly normalized BMN operators with, respectively, zero,
one and two impurities:
\begin{myeqnarray}
\cO^J \eq \frac{1}{\sqrt{J\, N^J}}\, \Tr \, Z^J ,\nn\\
\cO^{J}_{i} \eq \frac{1}{\sqrt{N^{J+1}}}\, \Tr \bigbrk{ \phi_i\, Z^J }
\label{single},\\
\cO^{J}_{ij,n} \eq \frac{1}{\sqrt{J \, N^{J+2}}}
\lrbrk{
\sum_{p=0}^J e^{2\pi i p\, n /J}\Tr \bigbrk{\phi_i\, Z^p\, \phi_j\, Z^{J-p}} 
-\delta_{ij}\Tr \bigbrk{\Zb Z^{J+1}} }.\nn 
\end{myeqnarray}
Recall that we shortened the notation by omitting the $x$-dependence 
of the operators and correlators,
and by setting all multiplicative factors $\frac{\gym^2}{8 \pi^2}$
to one. Both dependencies are trivially restored if needed,
\emph{cf} section 3.1.
The last operator has been slightly generalized as compared
to eq.\eqref{olddef} in order to also allow for the insertions
of two impurities of the same kind (i.e.~$i =j$). This was derived
in detail in \cite{Parnachev:2002kk}.  
As we just argued we should also include double-trace operators
(see also \cite{Bianchi:2002rw}). The ones we need 
(see \figref{fig:PP.TorusCut}) are defined as
\begin{myeqnarray}
\cT^{J,r} \eq  \cO^{r\cdot J} \, \cO^{(1-r)\cdot J}  ,
\nn\\
\cT^{J,r}_{i} \eq  \cO^{r\cdot J}_{i} \, \cO^{(1-r)\cdot J}  ,
\nn\\
\cT^{J,r}_{ij,n} \eq  \cO^{r\cdot J}_{ij,n}\, \cO^{(1-r)\cdot J} ,
\nn\\
\cT^{J,r}_{ij} \eq  \cO^{r\cdot J}_{i} \, \cO^{(1-r)\cdot J}_{j} . 
\label{double}
\end{myeqnarray}
Here $r=\frac{J'}{J}$ where $J'$ is taken in the range
$J'=1, \ldots, J-1$. In the BMN limit $r$ can be thought of as
a real number in the range $r \in (0,1)$.

The operators $\cO^J$ and $\cT^{J,r}$ are SO(4) singlets 
and the operators $\cO^J_{i}$ and $\cT^{J,r}_{i}$ are
SO(4) vectors. 
Operators containing two scalar defects should be decomposed into 
the SO(4) irreps $\mathbf{9}+\mathbf{6}+\mathbf{1}$.
They correspond to the symmetric-traceless,
anti-symmetric and singlet representations:
\begin{myeqnarray}
\cO^J_{(ij),n}\eq\half
\bigbrk{\cO^J_{ij,n}+\cO^J_{ji,n}}-\sfrac{1}{4}\,\delta_{ij} 
\tsum_k \cO^J_{kk,n}, 
\nn\\
\cO^J_{[ij],n}\eq\half
\bigbrk{\cO^J_{ij,n}-\cO^J_{ji,n}},\qquad\qquad
 \nn\\
\cO^J_{\singlet,n}\eq\half \tsum_k \cO^J_{kk,n}.
\label{decompose}
\end{myeqnarray}
Note that due to the identity $\cO^{J}_{ij,-n}=\cO^{J}_{ji,n}$
the operators 
in \eqref{decompose} with negative mode number $-n$
equal the operators with positive mode number $n$ 
up to a sign for the anti-symmetric operator
\begin{equation}
\cO^J_{(ij),-n}=\cO^J_{(ij),n},\quad
\cO^J_{[ij],-n}=-\cO^J_{[ij],n},\quad
\cO^J_{\singlet,-n}=\cO^J_{\singlet,n}.
\end{equation}
Clearly the zero-mode operators exist only in the
symmetric-traceless and singlet representations,
they are protected half BPS operators.
Nevertheless, 
we prefer not to implement the decomposition 
into symmetric and anti-symmetric parts in the course of our calculation:
This would lead to Fourier-sine and Fourier-cosine series instead of 
the more convenient ordinary Fourier series.
We therefore continue to work with $\cO^{J}_{12,n}$ 
to capture the anomalous dimension of the 
(redefined) $\mathbf{9}+\mathbf{6}$ 
operators (section \ref{ssec:2ptirrep})
according to 
\begin{equation}
\bigvev{\bar\cO^J_{12,n}\,\cO^J_{12,m}}
=\delta_{mn}
+\delta_{mn}\frac{\delta \Delta_n^{\mathbf{9}}+\delta \Delta_n^{\mathbf{6}}}{2}\,L
+\delta_{m,-n}\frac{\delta \Delta_n^{\mathbf{9}}-\delta \Delta_n^{\mathbf{6}}}{2}\,L.
\end{equation}
The singlet operator $\cO^J_{\singlet,n}$ needs to be
considered separately (section \ref{ssec:2ptsinglet}).
%
%
%
%

%
%

\subsection{Symmetric and anti-\-symmetric BMN operators}
\label{ssec:2ptirrep}

Next one computes the two-point functions of one- and two-trace operators
at the tree and one-loop level, up to ${\cal O}(\gtwo^2)$.
These computations are efficiently performed using the techniques of
appendices \ref{app:effver} or \ref{app:diagram}, which reduce the 
problem to a purely combinatorial
one. We already stated the known result for the overlap of
single-trace operators, eq.\eqref{startingpoint}. The analogous
expressions for double-trace operators are only needed to leading
order in $\gtwo$ and are therefore simpler:
\begin{myeqnarray}
\bigvev{\cT^{J,r}_{12}\, \bar\cT^{J,s}_{12}} \eq\delta_{rs}
+\order{\gtwo^2},
\nn\\
\bigvev{\cT^{J,r}_{12,m}\, \bar\cT^{J,s}_{12,n}} \eq
\delta_{rs}\delta_{mn}
\lrbrk{1+\frac{\lambda' L }{8\pi^2}\,\frac{8\pi^2m^2}{r^2}}
+\order{\gtwo^2},
\nn\\
\bigvev{ \cT^{J,r}_{12,m}\,\bar\cT^{J,s}_{12}}
 \eq \order{\gtwo^{2}}.
\label{doublecorrs}
\end{myeqnarray}
The overlaps between single- and double-trace
operators turn out  to be
\begin{myeqnarray}
\bigvev{\cT^{J,r}_{12,m}\, \bar\cO^J_{12,n}}
\eq\frac{\gtwo\,r^{3/2}\sqrt{1-r}\sin^2(\pi n r)}{\sqrt{J}\,\pi^2(m-n r)^2}
\nl\qquad\qquad
\times\lrbrk{1+\frac{\lambda'L}{8\pi^2}\,\frac{8\pi^2(m^2-mn r+n^2r^2)}{r^2}}
+\order{\gtwo^3},
\nn\\
\bigvev{\cT^{J,r}_{12}\, \bar\cO^J_{12,n}} \eq
\frac{\gtwo}{\sqrt{J}}
\lrbrk{\delta_{n,0}\, r-\frac{\sin^2(\pi n r)}{\pi^2\,n^2}}
\lrbrk{1+\frac{\lambda'L}{8\pi^2} \,8\pi^2 n^2}
+\order{\gtwo^3}.
\nn\\
\label{double-single}
\end{myeqnarray}
The classical parts were already found
in \cite{Constable:2002hw}
and the one-loop results for the special case $m=0$
were presented in \cite{Chu:2002pd}.
As we mentioned above
these overlaps are of $\order{\gtwo}$. 
To correctly diagonalize we have to 
remove the overlap by a redefinition of the single and double trace
operators
\footnote{
At $\order{\gtwo}$ the double trace
operator receives corrections from triple trace operators as
well. These, however, do not influence the anomalous
dimension of the single trace operators 
at $\order{\gtwo^2}$, since the overlap between a single trace
state and a triple trace state is itself already of ${\cal O}(g_2^2)$.}
\footnote{
The first of these equations was presented by S.Minwalla
at Strings 2002, along with a (tentative) result for the
anomalous dimensions. The latter disagrees with our findings.}
\begin{myeqnarray}
\cO^{\prime\, J}_{12,m}\eq
\cO^J_{12,m}
-\sum_{k,r}\frac{\gtwo\, r^{3/2}\sqrt{1-r}\,\sin^2(\pi m r)\, k}
{\sqrt{J}\,\pi^2(k-m r)^2(k+mr)}\,\cT^{J,r}_{12,k},
\nn\\
\cT^{\prime \,J,r}_{12,m}\eq
\cT^{J,r}_{12,m}
-\sum_k\frac{\gtwo\,r^{3/2}\sqrt{1-r}\,\sin^2(\pi kr)\,kr}{\sqrt{J}\,\pi^2 (m-kr)^2(m+kr)}
\,\cO^{J}_{12,k},
\nn\\
\cT^{\prime\, J,r}_{12}\eq
\cT^{J,r}_{12}
-\sum_k
\frac{\gtwo}{\sqrt{J}}
\lrbrk{\delta_{k,0}\,r-\frac{\sin^2(\pi k r)}{\pi^2 k^2}} 
\cO^{J}_{12,k}
\label{mix}
\end{myeqnarray}
where the sums go over all integers $k$ and
all $r=J'/J$ with $1\leq J'\leq J-1$; 
i.e.~the redefinitions are chosen such that
\begin{myeqnarray}
\bigvev{\cT^{\prime\, J,r}_{12,m}\, \bar\cO^{\prime\, J}_{12,n}}
\eq\order{\gtwo^3},
\nn\\
\bigvev{\cT^{\prime\, J,r}_{12}\, \bar\cO^{\prime\, J}_{12,n}} \eq
\order{\gtwo^3}.
\label{overlap}
\end{myeqnarray}
The eqs.\eqref{mix} are unique since 
we need to eliminate both the
leading  ${\cal O}(\gtwo)$ classical as well as the leading 
${\cal O}(\gtwo \lambda' L)$ one-loop overlaps, \emph{cf} 
eq.\eqref{double-single}.

The redefined single trace correlator receives corrections
from the double trace part
\begin{myeqnarray}
\lefteqn{\bigvev{\cO^{\prime\, J}_{12,m} \bar\cO^{\prime\, J}_{12,n}}}
\nle
\delta_{mn}
\lrbrk{1+\frac{\lambda' L }{8\pi^2} \,8\pi^2m^2}
+\gtwo^2\lrbrk{M_{mn}^1+\frac{\lambda' L}{8\pi^2}\lrbrk{8\pi^2 mn M^1_{mn}+\mathcal{D}^1_{mn}}}
\nl
-\gtwo^2\sum_{k,r}
\frac{r^{3}(1-r)\sin^2(\pi m r)\sin^2(\pi n r)\,k}
{J \pi^4(k-mr)^2(k+mr)(k-n r)^2(k+nr)}
\nl
\qquad\times\lrbrk{k+mr+nr+\frac{\lambda' L}
{8\pi^2r^2}\,8\pi^2\lrbrk{k^3+m^3r^3+n^3r^3}}
\nle
\delta_{mn}
\lrbrk{1+\frac{\lambda' L}{8\pi^2}\,8\pi^2m^2}
+\gtwo^2\lrbrk{M_{mn}+\frac{\lambda' L}{8\pi^2}\,\lrbrk{8\pi^2mn M_{mn}+\mathcal{D}_{mn}}}
\nn
\end{myeqnarray}
where we have defined
\begin{equation}
M_{mn}=M^1_{mn}+M^2_{mn},\qquad
\mathcal{D}_{mn}=\mathcal{D}^1_{mn}+\mathcal{D}^2_{mn},
\label{MD}
\end{equation}
and where the matrix elements $M^2_{mn}$, $\mathcal{D}^2_{mn}$
are derived by computing the above double sum in the BMN limit.
Their detailed form can be found in appendix \ref{app:matel}.
Care has to be taken to correctly treat the special cases $|m|=|n|$,
which are the ones that are directly relevant to the numerical
values of the toroidal anomalous dimensions.
The off-diagonal pieces are, however, needed as well.
They are part of the precise dictionary since
we have to work with an orthonormal set of 
operators in order to read off the one-loop correction to the
anomalous dimension, \emph{cf} the discussion around
eq.\eqref{extract}. We therefore linearly redefine the $\cO^{\prime\, J}_{12,n}$ 
once more, employing the off-diagonal elements:
\footnote{
At $\order{\gtwo^2}$ the single trace operators 
receive corrections from triple trace operators, 
but without effect for the anomalous dimensions at 
$\order{\gtwo^2}$.}
\begin{equation}
\cO^{\prime \prime\, J}_{12,m}=
\cO^{\prime\, J}_{12,m}+\gtwo^2 \,
\sum_k
T_{mk}\,\cO^{\prime\, J}_{12,k}
\label{jan1}
\end{equation}
with (here $|n| \neq |k|$)
\begin{equation}
T_{nn}=-\frac{M_{nn}}{2},\quad
T_{n,-n}=-\frac{M_{n,-n}}{2},\quad
T_{nk}=-\frac{n M_{nk}}{n+k}+\frac{\mathcal{D}_{nk}}{8\pi^2(n^2-k^2)}.
\label{jan2}
\end{equation}
This redefinition is chosen in order to remove the overlap
between impurity operators $\cO^{\prime\prime\, J}_{12,m}$ and $\bar\cO^{\prime\prime\, J}_{12,n}$
with $|n| \neq |m|$:
\begin{equation}
\bigvev{\cO^{\prime\prime\, J}_{12,m}\bar\cO^{\prime\prime\, J}_{12,n}}=\order{\gtwo^4}
\qquad \mbox{if } |m| \neq |n|
\end{equation} 
As was discussed above, a priori only
the symmetric and anti-symmetric combinations
$\cO^{\prime\prime\, J}_{(12),n}$ and $\cO^{\prime\prime\, J}_{[12],n}$ of eq.\eqref{decompose}
are expected to have definite anomalous dimensions; therefore
we should have $\bigvev{\cO^{\prime\prime\, J}_{12,n}\bar\cO^{\prime \prime\, J}_{12,-n}} \neq 0$.
Surprisingly, we find, using eq.\eqref{MD} and the results of
appendix \ref{app:matel},
\begin{equation}
\frac{\delta\Delta^{\mathbf{9}}_n-\delta\Delta^{\mathbf{6}}_n}{2}\,L
=
\bigvev{\cO^{\prime \prime\, J}_{12,n}\bar\cO^{\prime \prime\, J}_{12,-n}}= 
\frac{\gtwo^2\lambda'}{8\pi^2}~
(\mathcal{D}_{n,-n} -16 \pi^2 n^2 M_{n,-n})L=0.
\end{equation}
This involves a delicate conspiracy between the matrix elements 
$M_{n,-n}$ and $\mathcal{D}_{n,-n}$,
which are, respectively, obtained from rather different calculations.
It means that $\cO^{\prime \prime\, J}_{(12),n}$ and $\cO^{\prime \prime\, J}_{[12],n}$ have
\emph{degenerate} anomalous dimensions, and the
operators $\cO^{\prime \prime\, J}_{12,n}$ are completely one-loop orthogonal
up to order $\gtwo^2$
\begin{myeqnarray}
\bigvev{\cO^{\prime \prime\, J}_{12,m}\bar\cO^{\prime \prime\, J}_{12,n}}
\eq\delta_{mn} \lrbrk{1+
\frac{\lambda' L}{8\pi^2}\lrbrk{8\pi^2m^2
+\gtwo^2~\mathcal{D}_{m m}}}  
\nle
\delta_{mn} \Bigg( 1+
\frac{\lambda' L}{8\pi^2}\bigg(8\pi^2m^2
+\gtwo^2 \Big( \frac{1}{6}+\frac{35}{16\pi^2m^2}
\Big) \bigg) \Bigg),
\end{myeqnarray}
(we used $\mathcal{D}_{mm}=\mathcal{D}^1_{mm}+\mathcal{D}^2_{mm}$
and appendix \ref{app:matel}) 
allowing us, in view of eqs.\eqref{defdeltaDelta},\eqref{extract2},
to read off their anomalous dimensions
\begin{equation}
\Delta^{\mathbf{9}}_n=\Delta^{\mathbf{6}}_n=J+2+
\frac{\lambda'}{8\pi^2}\lrbrk{8\pi^2 n^2+\gtwo^2 
\lrbrk{\frac{1}{6}+\frac{35}{16\pi^2n^2}}
} \quad (n \neq 0).
\label{dim}
\end{equation}
Put differently, operators in the SO(4) representations 
$\mathbf{9}$ and $\mathbf{6}$ possess degenerate anomalous dimensions.
{}From the field theory point of view there is a priori no 
reason to believe that operators belonging to different 
representations should have equal anomalous dimensions. 
On the sphere it might have been a coincidence 
that the dimensions match, but on the torus it 
is a remarkable result. It would be interesting to
understand the symmetry reason for this result.

\subsection{Singlet BMN operators}
\label{ssec:2ptsinglet}

Let us now turn to the determination of the anomalous
dimension of the SO(4) singlet BMN operator with
two impurities. Again the mixing of one- and two-trace
operators needs to be taken
into account. Here the computations are somewhat more 
involved as contributions from the ``K-terms'' of
\eqref{master} coupling to traces need to be dealt with,
see appendices \ref{app:effver} and \ref{app:diagram} for details. Note, that now the inclusion
of the $\Tr(\bar Z Z^{J+1})$ term for the diagonal operator 
${\cal O}_{ii,n}^J$ of \eqref{single} is crucial:
It precisely cancels terms violating the BMN scaling limit
originating from the first ``naive'' piece of the
operator ${\cal O}_{ii,n}^J$ in \eqref{single}. We then find
\begin{myeqnarray}
\bigvev{\cO^J_{\singlet,m}\, \bar\cO^J_{\singlet,n}} \eq
\bigvev{\cO^J_{12,m}\, \bar\cO^J_{12,n}}
+\bigvev{\cO^J_{12,m}\, \bar\cO^J_{12,-n}}
-\frac{\lambda' L}{8\pi^2}\, 2 \gtwo^2\mathcal{D}^1_{mn}
+\order{\gtwo^4},
\nn\\
\bigvev{\cT^{J,r}_{\singlet,m}\, \bar\cO^J_{\singlet,n}} \eq
\bigvev{\cT^{J,r}_{12,m}\, \bar\cO^J_{12,n}}
+\bigvev{\cT^{J,r}_{12,m}\, \bar\cO^J_{12,-n}}
\nl
-\,\frac{\lambda' L}{8\pi^2}\,
\frac{16\gtwo \sqrt{1-r}\sin^2(\pi n r)}{\sqrt{Jr}}
+\order{\gtwo^3},
\nn\\
\bigvev{\cT^{J,r}_{\singlet}\, \bar\cO^J_{\singlet,n}} \eq
2\, \bigvev{\cT^{J,r}_{12}\, \bar\cO^J_{12,n}}
+ \frac{\lambda'L}{8\pi^2}\, \frac{16 \gtwo\sin^2(\pi n r)}{\sqrt{J}}
+\order{\gtwo^3},
\nn\\
\bigvev{\cT^{J,r}_{\singlet}\, \bar\cT^{J,s}_{\singlet}} \eq
\bigvev{\cT^{J,r}_{12}\, \bar\cT^{J,s}_{12}}
+\bigvev{\cT^{J,r}_{12}\, \bar\cT^{J,1-s}_{12}}
+\order{\gtwo^2},
\nn\\
\bigvev{\cT^{J,r}_{\singlet,m}\, \bar\cT^{J,s}_{\singlet,n}}\eq
\bigvev{\cT^{J,r}_{12,m}\, \bar\cT^{J,s}_{12,n}}
+\bigvev{\cT^{J,r}_{12,m}\, \bar\cT^{J,s}_{12,-n}}
+\order{\gtwo^2},
\nn\\
\bigvev{ \cT^{J,r}_{\singlet,m}\,\bar\cT^{K,s}_{\singlet}} \eq
\order{\gtwo^{2}}.
\label{singcorrs}
\end{myeqnarray}
where the contributions from $V_F$ arising from  \eqref{master} reside
in the correlator expressions in the right hand sides of the above
and the contributions from the $V_K$ sector have been
spelled out explicitly to the order needed. Making use of 
\eqref{startingpoint},\eqref{doublecorrs} and \eqref{double-single}
one has
\begin{myeqnarray}
\bigvev{\cO^J_{\singlet,m}\, \bar\cO^J_{\singlet,n}} \eq
\bigbrk{\delta_{mn}+\delta_{m,-n}}
\lrbrk{1+\frac{\lambda'L}{8\pi^2}\,8\pi^2 m^2}
\nl
+\gtwo^2\,\lrbrk{M^1_{mn}+M^1_{m,-n}
+\frac{\lambda' L}{8\pi^2} 8\pi^2mn \lrbrk{M^1_{mn}-M^1_{m,-n}}}
\nl
+\order{\gtwo^4},
\nn\\
\bigvev{\cT^{J,r}_{\singlet,m}\, \bar\cO^J_{\singlet,n}} \eq
\frac{\gtwo\,r^{3/2}\sqrt{1-r}\sin^2(\pi n r)}{\sqrt{J}\,\pi^2(m-n r)^2}
\lrbrk{1+\frac{\lambda'L}{8\pi^2}\,\frac{8\pi^2 mn}{r}}
\nl
+\frac{\gtwo\,r^{3/2}\sqrt{1-r}\sin^2(\pi n r)}{\sqrt{J}\,\pi^2(m+n r)^2}
\lrbrk{1-\frac{\lambda'L}{8\pi^2}\,\frac{8\pi^2 mn}{r}}
+\order{\gtwo^3},
\nn\\
\bigvev{\cT^{J,r}_{\singlet}\, \bar\cO^J_{\singlet,n}} \eq
\frac{2\gtwo}{\sqrt{J}}
\lrbrk{\delta_{n,0}\, r -\frac{\sin^2(\pi n r)}{\pi^2\,n^2}}
+\order{\gtwo^3},
\nn\\
%
\bigvev{\cT^{J,r}_{\singlet}\, \bar\cT^{J,s}_{\singlet}} \eq
\delta_{rs}+\delta_{r,1-s}
+\order{\gtwo^2},
\nn\\
\bigvev{\cT^{J,r}_{\singlet,m}\, \bar\cT^{J,s}_{\singlet,n}} \eq
\delta_{rs}\bigbrk{\delta_{mn}+\delta_{m,-n}}
\lrbrk{1+\frac{\lambda' L}{8\pi^2}\,\frac{8\pi^2 m^2}{r^2}}
+\order{\gtwo^2},
\nn\\
\bigvev{ \cT^{J,r}_{\singlet,m}\,\bar\cT^{J,s}_{\singlet}} \eq
\order{\gtwo^{2}}.
\end{myeqnarray}
Again to correctly diagonalize the singlet operators we need
to remove the overlap with double trace operators
by a redefinition through
\begin{myeqnarray}
\cO^{\prime\, J}_{\singlet,m}\eq
\cO^J_{\singlet,m}
-\sum_{k,r}\frac{\gtwo\, r^{3/2}\sqrt{1-r}\,\sin^2(\pi m r)\, mr}
{\sqrt{J}\,\pi^2(k-m r)^2(k+mr)}\,\cT^{J,r}_{\singlet,k}
\nl
-\sum_r \frac{\gtwo}{\sqrt{J}}\lrbrk{ \delta_{m,0}r -\frac{\sin^2 (\pi m r)}{\pi^2 m^2}  }
\,\cT^{J,r}_{\singlet},
\nn\\
\cT^{\prime\, J,r}_{\singlet,m}\eq
\cT^{J,r}_{\singlet,m}
-\sum_k\frac{\gtwo\,r^{3/2}\sqrt{1-r}\,\sin^2(\pi kr)\,m}{\sqrt{J}\,\pi^2 (m-kr)^2(m+kr)}
\,\cO^{J}_{\singlet,k},
\nn\\
\cT^{\prime\, J,r}_{\singlet}\eq\cT^{J,r}_{\singlet}\, .
\label{singlemix}
\end{myeqnarray}
This redefinition is such that
\begin{myeqnarray}
\bigvev{\cT^{\prime\, J,r}_{\singlet,m}\, \bar\cO^{\prime \,J}_{\singlet,n}}
\eq\order{\gtwo^3},
\nn\\
\bigvev{\cT^{\prime\, J,r}_{\singlet}\, \bar\cO^{\prime\, J}_{\singlet,n}} \eq 
\order{\gtwo^3}
\label{singoverlap}
\end{myeqnarray}
as before. Proceeding, one obtains for the modified
single trace correlator (where $m\neq 0 \neq n$)
\begin{myeqnarray}
\lefteqn{ \bigvev{\cO^{\prime\, J}_{\singlet,m}\, \bar\cO^{\prime\, J}_{\singlet,n}} }
\nle
\bigbrk{\delta_{mn}+\delta_{m,-n}}
\lrbrk{1+\frac{\lambda'L}{8\pi^2}\,8\pi^2 m^2}
\nl
+\gtwo^2\,\lrbrk{M^1_{mn}+M^1_{m,-n}
+\frac{\lambda' L}{8\pi^2} 8\pi^2mn \lrbrk{M^1_{mn}-M^1_{m,-n}}}
\nl
-\sum_{k,r}
\frac{\gtwo^2\, r^3(1-r)\,\sin^2(\pi m r)\sin^2(\pi n r)\, }
{J\,\pi^4(k-m r)^2(k+mr)(k-n r)^2(k+nr)}
\nl
\qquad\times\lrbrk{kmr+knr+nmr^2+\frac{\lambda'L}{8\pi^2}\,8\pi^2 kmn(k+mr+nr)}
\nl
-\sum_{k,r}
\frac{\gtwo^2\, r^3(1-r)\,\sin^2(\pi m r)\sin^2(\pi n r)\, }
{J\,\pi^4(k-m r)^2(k+mr)(k+n r)^2(k-nr)}
\nl
\qquad\times\lrbrk{kmr-knr-nmr^2-\frac{\lambda'L}{8\pi^2}\,8\pi^2 kmn(k+mr-nr)}
\nl
-\sum_r \frac{2\gtwo^2\,\sin^2(\pi m r)\, \sin^2(\pi n r)}{J\pi^4 m^2 n^2}
\nle
\bigbrk{\delta_{mn}+\delta_{m,-n}}
\lrbrk{1+\frac{\lambda'L}{8\pi^2}\,8\pi^2 m^2}
\nl
+\gtwo^2\,\lrbrk{M'_{mn}+M'_{m,-n}
+\frac{\lambda' L}{8\pi^2} 8\pi^2mn \lrbrk{M'_{mn}-M'_{m,-n}}}
\end{myeqnarray}
where we have defined
\begin{equation}
M'_{mn}=M^1_{mn}+M^3_{mn}
=-\half M^1_{mn}.
\end{equation}
%
The matrix elements $M^3_{mn}=-\frac{3}{2}M^1_{mn}$ 
are derived by evaluating the above sums. 
Curiously, here there is no analogue of the matrix
${\cal D}_{mn}$ appearing. Just as in the discussion of the
previous subsection on the symmetric and anti-symmetric
BMN operators another linear redefinition of singlet operator
${\cal O}^{\prime\, J}_{\singlet,n}$ is needed in order to read off
the one-loop correction to the anomalous dimension:
\begin{equation}
\cO^{\prime \prime\, J}_{\singlet,m}=
\cO^{\prime\, J}_{\singlet,m}+\gtwo^2 \,
\sum_k
T'_{mk}\,\cO^{\prime\, J}_{\singlet,k}
\label{single''}
\end{equation}
with (here $|n| \neq |k|$)
\begin{equation}
T'_{nn}=-\frac{M'_{nn}}{2},\quad
T'_{n,-n}=-\frac{M'_{n,-n}}{2},\quad
T'_{nk}=-\frac{n M'_{nk}}{n+k}
\end{equation}
in great similarity to \eqref{jan2}.
This second redefinition removes the overlap
between singlet operators $\cO^{\prime \prime\, J}_{\singlet,m}$ and $\bar\cO^{\prime \prime\, J}_{\singlet,n}$
with $|n| \neq |m|$:
\begin{equation}
\bigvev{\cO^{\prime \prime\, J}_{\singlet,m}\bar\cO^{\prime \prime\, J}_{\singlet,n}}=\order{\gtwo^4}
\qquad \mbox{if } |m| \neq |n|
\end{equation} 
Turning to the case $|n| = |m|$ one finds
\begin{equation}
\bigvev{\cO^{\prime \prime\, J}_{\singlet,|n|}\bar\cO^{\prime \prime\, J}_{\singlet,|n|}}=
1+\lambda'\, L \Bigl ( n^2 -2 \gtwo^2\, n^2\, M'_{n,-n}
+\order{\gtwo^4}\, \Bigr )
\end{equation}
which upon making use of the formulas of appendix \ref{app:matel} for
$M'_{n,-n}=-\half M^1_{n,-n}$ 
leads to the surprising
result ($n,m \neq 0$)
\begin{equation}
\bigvev{\cO^{\prime \prime\, J}_{\singlet,m}\bar\cO^{\prime \prime\, J}_{\singlet,n}}=
(\delta_{m,n}+\delta_{m,-n})\,\Bigg( 1+
\frac{\lambda' L}{8\pi^2}\bigg(8\pi^2m^2
+\gtwo^2 \Big( \frac{1}{6}+\frac{35}{16\pi^2m^2}
\Big) \bigg) \Bigg)
\label{singletad}
\end{equation}
manifesting our claim that the singlet BMN operators carry
the same anomalous dimension 
$\Delta^{\mathbf{1}}_n=\Delta^{\mathbf{6}}_n=\Delta^{\mathbf{9}}_n$
of \eqref{dim} as the BMN operators in the 
SO(4) representations $\mathbf{9}$ and $\mathbf{6}$.
Note that the delta-function structure 
in \eqref{singletad} originates from the identity
$\cO^{\prime \prime\, J}_{\singlet,n}=\cO^{\prime \prime\, J}_{\singlet,-n}$.
This observed toroidal degeneracy of all two impurity 
SO(4) BMN operators is rather remarkable. It would be very
desirable to understand it from the dual string perspective.

\subsection{Operator mixing for chiral primaries}
\label{ssec:2ptlevelzero}

The redefinition of the original BMN impurity operators
eqs.\eqref{mix} is uniquely determined by demanding that the overlap
eqs.\eqref{overlap} between redefined single and double trace
operators vanishes to order $\gtwo$. For protected operators,
such as $\cO^J$, $\cO^{J}_{i}$ and $\cO^{J}_{ij,0}$, the 
one-loop correction vanishes automatically.
Therefore, there are much less constraints on the  
operator mixing. It thus seems that the dictionary relating pp-strings
and gauge theory is highly non-unique as far as massless string modes and
the corresponding protected operators are concerned. This 
appears to render string-scattering
involving ``graviton'' states $|0,p^+ \rangle$ 
ambiguous. It is not clear to us how to fix the large
freedom in defining the constants $a_{J,r}$,$b_{J,r}$: 
\begin{myeqnarray}
|0,p^+ \rangle \leftrightarrow 
\cO^{\prime\, J}\eq
\cO^J -\frac{\gtwo}{\sqrt{J}} \sum_r a_{J,r}~\cT^{J,r} + {\cal O}(\gtwo^2),
\nn\\[0.2cm]
|0,p^+_1 \rangle \otimes |0,p^+_2 \rangle \leftrightarrow 
\cT^{\prime\, J,r}\eq
\cT^{J,r}-\frac{\gtwo}{\sqrt{J}}~b_{J,r}~\cO^{J}+{\cal O}(\gtwo^2).
\label{chiralmix}
\end{myeqnarray}
Given arbitrary constants $a_{J,r}$, we can always solve for
$b_{J,r}$ in order to satisfy
$\bigvev{\cT^{\prime\, J,r}\, \bar\cO^{\prime\, J}} = \order{\gtwo^3}$. 
In fact, for every set of operators with equal
scaling dimensions and quantum numbers
the freedom to redefine the operators by an
orthogonal transformation remains. Here, the
freedom is manifested in the undetermined parameters $a_{J,r}$.

\subsection{Further comments}
\label{ssec:2ptcomment}

Clearly numerous extensions of the above calculations are
possible, if tedious. 
In particular, it would be extremely interesting to use our 
effective vertex procedure and continue the above one-loop 
diagonalization to higher genus. As was discussed already in the
introduction, see discussion surrounding eq.\eqref{true}, 
if it is true \cite{Constable:2002hw,Verlinde:2002ig} that string
interactions have to be identified with $\sqrt{\lambda'} \gtwo$ as 
opposed to just $\gtwo$, \emph{we should find that} eq.\eqref{dim} 
\emph{is the exact one-loop anomalous dimension to all orders in
$\frac{1}{N^2}$}! The double torus, i.e.~${\cal O}(\gtwo^4)$ should 
only contribute at the two-loop (${\cal O}(\lambda^{\prime\, 2})$) level.
{}From the point of view of the gauge theory this would be a miracle.
Two- and higher loop calculations are also desirable; it would be 
very interesting to work out the effective vertices for these
cases and investigate whether a simple all-orders pattern exists.

\sect{Three-point functions and BMN operator mixing}
\label{sec:3pt}

\subsection{General remarks}
\label{ssec:3ptintro}

In chapter \ref{sec:4pt} we analyzed four-point functions in the BMN limit
and found them to be affected by two kinds of pathologies:
Space-time discontinuities at the classical level, and bad
large $J$ scaling at the quantum level. It would be very
interesting if a procedure could be found that renders 
them meaningful and/or allows them to become part of the
gauge theory-string dictionary. On a technical level, these
results are maybe not too surprising. Clearly these pathologies
are intimately related, respectively, to the fact that the 
space-time form of four point functions is \emph{not} determined
by conformal invariance, and to the fact that non-protected
operators appear in their double-operator product expansion
(see e.g.~\cite{Bianchi:2002rw,Arutyunov:2002rs} and references
therein). But these explanations immediately suggest that 
\emph{three}-point functions might nevertheless be consistent
in the BMN limit: On the one hand, their space-time structure is 
fixed, and on the other no unprotected fields appear in
intermediate channels. And indeed one finds that the above pathologies
are not present for various classical and quantum calculations
involving BMN three-point functions 
\cite{Kristjansen:2002bb,Constable:2002hw,Chu:2002pd}.
However, it turns out that a different pathology nevertheless affects the
one-loop quantum corrections of three-point functions of impurity BMN
operators. Due to the conformal symmetry, a three point function
of conformal operators ${\cal O}_i(x)$ with scaling dimensions
$\Delta_i$ has to be of the form
\begin{equation}
\bigvev{\cO_i(x_1) \cO_j(x_2) \cO_k(x_3)}=
C_{i j k}~F^{\Delta_1\Delta_2\Delta_3}_{x_1x_2x_3},
\label{Cdef}
\end{equation}
where ($x_{ij}=x_i-x_j$)
\begin{equation}
F^{\Delta_1\Delta_2\Delta_3}_{x_1x_2x_3}
=\frac{1}{|x_{12}|^{\Delta_1+\Delta_2-\Delta_3}\,
|x_{23}|^{\Delta_2+\Delta_3-\Delta_1}\,
|x_{31}|^{\Delta_3+\Delta_1-\Delta_2}}.
\label{Fdef}
\end{equation}
If one computes the one-loop contribution to 
$\bigvev{\bar \cO^{J}_{ij,n}(x_1) 
\cO^{r\cdot J}_{ij,m}(y_1)\cO^{(1-r)\cdot J}(y_2)}$
for e.g.~the two-impurity operators in eq.\eqref{single} for 
$m,n \neq 0$ one finds 
(see footnote \ref{fn:Chu} on page \pageref{fn:Chu})
that the result cannot be brought into the form of 
eqs.\eqref{Cdef},\eqref{Fdef} (in the special case of $m=0$ this
problem does not occur, as has been shown in \cite{Chu:2002pd}).
This puzzle has a beautiful resolution, as will be shown in the next
section. Three-point functions are down by one factor of $\frac{1}{N}$ 
w.r.t.~two-point functions. Considering the operator mixing 
equations eqs.\eqref{mix},\eqref{singlemix} we see that the
three-point functions receive corrections: Each single trace operator 
inside a three-point correlator is modified at 
${\cal O}(\frac{1}{N})$ by a double trace operator. The latter
potentially can, due to large $N$ factorization, combine with the
remaining single trace operators to give an overall ${\cal O}(\frac{1}{N})$
contribution, which is therefore actually of equal importance as
compared to the bare (i.e.~before mixing) correlator. 
This effect \emph{restores} conformal invariance. It means, once again,
that the gauge theory, quite independent from the requirements
imposed by building a pp-string dictionary, imposes on us the operator
mixing eqs.\eqref{mix},\eqref{singlemix} discussed previously.

However, the most suprising result found below is that even
the structure constants $C_{ijk}$ are modified,
and no longer agree with the classical three-point functions of
the original BMN operators in eq.\eqref{single}, as originally worked out
in \cite{Constable:2002hw}, section 3.2. As a consequence, some
doubt is cast onto the proposal of \cite{Constable:2002hw} which 
relates the string-field theory light-cone interaction vertex
to the $C_{ijk}$ (\emph{cf} eq.(5.4) in \cite{Constable:2002hw}). 
We feel, in support of the ideas presented in \cite{Verlinde:2002ig}, 
that it is an open question whether the gauge theory three-point functions 
will become part of the BMN dictionary.

\subsection{Three-point functions of redefined BMN operators}
\label{ssec:3ptfunction}

Let us then compute the three-point function of the redefined,
diagonalized BMN operators of eqs.\eqref{jan1},\eqref{single''},
up to one-loop and at the leading (${\cal O}(\gtwo)$) order in 
topology. {}From eqs.\eqref{Cdef},\eqref{Fdef} we expect 
%
\begin{myeqnarray}
\bigvev{\bar \cO^{\prime\prime\,J}_{ij,n}(x)\, \cO^{\prime\prime\,r\cdot J}_{kl,m}(y_1)\,\cO^{(1-r)\cdot J}(y_2)}
\eq C^{J,r}_{ij,n;kl,m}\,F^{\Delta^J_n\Delta^{rJ}_{m}\Delta^{(1-r)J}}_{xy_1y_2},
\nn\\
\bigvev{\bar \cO^{\prime\prime\, J}_{ij,n}(x)\, \cO^{r\cdot J}_{k}(y_1)\,\cO^{(1-r)\cdot J}_{l} (y_2)}
\eq C^{J,r}_{ij,n;kl}\,F^{\Delta^J_n\Delta^{rJ+1}\Delta^{(1-r)J+1}}_{xy_1y_2},
\end{myeqnarray}
where $F$ contains the space-time dependence.
The one-loop conformal scaling dimensions on the sphere
are $\Delta^J=J$, $\Delta^{rJ}_n=J+2+\lambda' n^2/r^2$.
We again decompose the correlators into the $\mathbf{9}$, $\mathbf{6}$
and $\mathbf{1}$ parts. Actually only the double trace correction
to the barred operators $\bar \cO^{\prime \prime \,J}_{ij,n}$ contributes.
One finds
\begin{myeqnarray}\label{eq:fourpointfour}
C_{ij,n;kl,m}^{J,r}\eq
\frac{2 g_2 \sqrt{1-r}\,\sin^2(\pi n r)}{\sqrt{Jr}\, \pi^2 (n^2-m^2/r^2)^2}
\lrbrk{1-\frac{\lambda' \lrbrk{n^2-m^2/r^2}}{2}}
\nl
\qquad\lrbrk{\delta_{i(k}\delta_{l)j}\,\,n^2+\delta_{i[k}\delta_{l]j}\,\frac{nm}{r}+
\sfrac{1}{4}\delta_{ij}\delta_{kl}\,\frac{m^2}{r^2}},
\nn\\
C_{ij,n;kl}^{J,r}\eq
\frac{2g_2}{\sqrt{J}}\lrbrk{\delta_{n,0}\, r
-\frac{\sin^2(\pi n r)}{\pi^2 n^2}}
\lrbrk{1-\frac{\lambda' n^2}{2}}
\delta_{i(k}\delta_{l)j}.
\end{myeqnarray}
%
%
%
The classical calculation proceeds by the technique employed
previously, and the quantum correction requires an analysis 
similar to the one in appendices \ref{app:4pt}, \ref{app:effver}
and \ref{app:diagram}. As we stressed above,
it is reassuring that the quantum correction is consistent with
the space-time structure imposed by conformal invariance.
As we claimed above, these structure constants differ from
the ones for the original BMN operators, \emph{cf} 
eqs.(3.10),(3.11) in \cite{Constable:2002hw}.\bigskip

\noindent
  \textbf{Note added in proof:} The $\lambda'$ contribution to
  the structure constants in \eqref{eq:fourpointfour} 
  actually cannot be fixed
  at this point due to possible redefinitions of the
  operators proportional to $\lambda'$.
  In the same way as operator mixing at one-loop modifies the 
  leading $\order{\lambda^{\prime\,0}}$ result,
  mixing at two-loops may modify \eqref{eq:fourpointfour} 
  at $\order{\lambda'}$.




\subsection*{Acknowledgments}

We would like to thank Gleb Arutyunov, Massimo Bianchi, Chong-Sun Chu,
Dan Freedman, Jakob Langgaard Nielsen,
and Herman Verlinde for interesting discussions. 
C. Kristjansen acknowledges the support of the EU network on
``Discrete Random Geometry'', grant HPRN-CT-1999-00161.

\appendix

\sect{Four-point functions at one-loop}
\label{app:4pt}

In this section we treat the computation of four-point functions 
of operators ${\cal O}^J(x)=\Tr Z^J(x)$ up to one-loop order.
Calculations for $J=2$ have been performed previously in
\cite{Gonzalez-Rey:1998tk,Eden:1998hh,Eden:1999kh,Bianchi:1999ge}.
First, we introduce and discuss some functions 
that play a central role in the computations. Next, we
present the correlators of the fields $Z(x)$ 
and finally we show how to 
construct from these the correlators of the operators ${\cal O}^J(x)$.

\subsection{Notation}  
\label{ssec:notation}

The field content of $\superN=4$ SYM in four dimensions  are the scalars, 
$\phi_i(x)$, $i\in \{1,\ldots,6\}$ which transform under the R-symmetry
group SO(6),  $A_\mu(x)$ with $\mu\in\{1,\ldots,4\}$ which is a space-time
vector and $\psi(x)$ which is a sixteen component spinor.
These fields are Hermitean $N\times N$ matrices and can be expanded
in terms of the generators $T^a$ of the gauge group U($N$) as
\begin{equation}
\phi_i(x)=\sum_{a=0}^{N^2-1} \phi^{(a)}_i(x)T^a,\quad
A_\mu(x)=\sum_{a=0}^{N^2-1} A_\mu^{(a)}(x)T^a ,\quad
\psi(x)=\sum_{a=0}^{N^2-1} \psi^{(a)}(x)T^a.
\end{equation}
The conventions for the generators are
\begin{equation}
\Tr (T^a T^b)=\delta^{ab},\quad
\sum_{a=0}^{N^2-1} (T^a)^\alpha{}_\beta(T^a)^\gamma{}_\delta=\delta^\alpha_\delta\delta^\gamma_\beta.
\end{equation}
Our Euclidean action of $\superN=4$ supersymmetric Yang-Mills
theory reads
\begin{myeqnarray}
\label{symaction}
S\eq\frac{2}{\gym^2}\int d^4x 
\Tr \Big(
\sfrac{1}{4}F_{\mu\nu}F_{\mu\nu}
+\sfrac{1}{2}D_\mu\phi_i D_\mu\phi_i
-\sfrac{1}{4}[\phi_i,\phi_j][\phi_i,\phi_j] +
\nl\qquad\qquad\qquad\qquad
+\sfrac{1}{2}\bar\psi\Gamma_\mu D_\mu\psi
-\sfrac{i}{2} \bar\psi \Gamma_i [\phi_i,\psi]
\Big)
\end{myeqnarray}
where $F_{\mu\nu}=\partial_\mu A_\nu-\partial_\nu A_\mu -i [A_\mu,A_\nu]$ 
and the covariant derivative is
$D_\mu\phi= \partial_\mu\phi_i-i[A_\mu,\phi_i]$. Furthermore,
$(\Gamma_\mu,\Gamma_i)$ are the ten-dimensional Dirac matrices in
the Majorana-Weyl representation. All our computation are done
in the Feynman gauge.

\subsection{Some functions}
\label{ssec:functions}

We introduce the scalar propagator and 
some fundamental tree functions
\begin{myeqnarray}\label{eq:PP.FundFun}
I_{12}\eq\frac{1}{(2\pi)^2(x_1-x_2)^2},
\nln
Y_{123}\eq\int d^4 w \,I_{1w} I_{2w} I_{3w},
\nln
X_{1234}\eq\int d^4 w \,I_{1w} I_{2w} I_{3w} I_{4w},
\nln
H_{12,34}\eq\int d^4 u \,d^4 v\, I_{1u} I_{2u} I_{uv} I_{3v} I_{4v}.
\end{myeqnarray}
We have put the space-time points as indices to the function
to make the expressions more compact. These functions are
all finite except in certain limits. For example $Y$, $X$ and $H$ 
diverge logarithmically when $x_1\to x_2$. 
The functions $X$ and $Y$ can be evaluated explicitly \cite{Usyukina:1993jd}
\begin{myeqnarray}
X_{1234}
\eq\mathrel{}\frac{\pi^2\Phi(r,s)}{(2\pi)^8 (x_1-x_3)^2(x_2-x_4)^2},
\nl
\qquad r=\frac{(x_1-x_2)^2(x_3-x_4)^2}{(x_1-x_3)^2(x_2-x_4)^2},\quad
s=\frac{(x_2-x_3)^2(x_4-x_1)^2}{(x_1-x_3)^2(x_2-x_4)^2},
\nln
Y_{123}\eq \lim_{x_4\to\infty} (2\pi)^2 x_4^2 X_{1234}.
\end{myeqnarray}
In the euclidean region 
($\sqrt{r}+\sqrt{s}\geq 1$, $\abs{\sqrt{r}-\sqrt{s}}\leq 1$) 
$\Phi$ can be written in a manifestly real fashion as
\begin{myeqnarray}
\Phi(r,s)\eq
\frac{1}{A}\Im\lrbrk{\Li_2 \frac{e^{i\varphi} \sqrt{r}}{\sqrt{s}}+
\ln\frac{\sqrt{r}}{\sqrt{s}}\,\ln\frac{\sqrt{s}-e^{i\varphi} \sqrt{r}}{\sqrt{s}}}
\nle
\frac{\abs{\ln r}}{\sqrt{s}}+\frac{\abs{\ln s}}{\sqrt{r}}\quad
(\mbox{for }
\abs{\sqrt{r}\pm\sqrt{s}}=1),
\nln
e^{i\varphi}\eq i\sqrt{-\frac{1-r-s-4i A}{1-r-s+4i A}},
\qquad
A=\quarter\sqrt{4rs-(1-r-s)^2}.
\end{myeqnarray}
It is positive everywhere, vanishes only in the limit
$r,s\to \infty$ and
has the hidden symmetry $\Phi(r,s)=\Phi (1/r,s/r)/r$.
The combinations $A$ and $\varphi$ can be interpreted geometrically:
By a conformal transformation move the point $x_4$ of $X$ to infinity
and scale such that $\abs{x_1-x_3}=1$.
The points $x_1,x_2,x_3$ span a triangle 
with area $A$ and angle $\varphi$ at $x_2$.

There seems to be no analytic expression for the function $H$, yet.
However, we need it only in the combination
\begin{myeqnarray}\label{eq:PP.GluExSimp}
F_{12,34}\eq
\frac{(\partial_1-\partial_2)\cdott (\partial_3-\partial_4) H_{12,34}}{I_{12}I_{34}}
\nle
\frac{X_{1234}}{I_{13}I_{24}}
-\frac{X_{1234}}{I_{14}I_{23}}
+G_{1,34}-G_{2,34}+G_{3,12}-G_{4,12},
\nln
G_{1,34}\eq\frac{Y_{134}}{I_{14}}-\frac{Y_{134}}{I_{13}}.
\end{myeqnarray}
The equality of the two expressions can be shown 
by transforming them to
momentum space and writing the inverse propagators as
derivatives on the momenta, i.e.
$1/I_{12}=(2\pi)^2(x_1-x_2)^2=-(2\pi)^2(\partial_{p_1}-\partial_{p_1})^2$.

\subsection{Scalar correlators}
\label{ssec:symcorr}

At one-loop we need radiative corrections to the 
propagators and $4$-point connected Green functions. 
These are the scalar self-energy, gluon exchange and 
scalar potential interactions.
The scalar propagator with self-energy corrections can be written as
\begin{myeqnarray}\label{eq:PP.PropG2}
\lefteqn{\bigvev{Z^a(x_1) \bar Z^b(x_2)}}
\nle
\half \gym^2 I_{12}
\lrbrk{\Tr T^a T^b-\gym^2 (N\Tr T^a T^b-\Tr T^a \Tr T^b)\frac{Y_{112}+Y_{122}}{I_{12}}}.
\qquad
\end{myeqnarray}
Note, that the momentum space representation 
$\int d^4 k/(2\pi)^4 p^2 k^2(p-k)^2$
of $Y_{112}$
is just the sum of contributions from the
scalar-gluon loop and the scalar tadpole.
The function $Y_{123}$ diverges logarithmically when
two of the points approach each other and 
$Y_{112}$ thus contains a logarithmic infinity.
The connected four-point function can be written as
multiplicative corrections to the free, disconnected 
propagators
\begin{myeqnarray}\label{eq:PP.4ptG2}
\lefteqn{\bigvev{Z^a(x_1) Z^b(x_2) \bar Z^c(x_3) \bar Z^d(x_4)}\indup{conn}}
\nle
\sfrac{1}{4} \gym^4 I_{13}I_{24}
\times(-\half \gym^2) \Tr [T^a,T^c][T^b,T^d]
\lrbrk{\frac{X_{1234}}{I_{13}I_{24}}+F_{13,24}}
\nl
+\sfrac{1}{4} \gym^4 I_{14}I_{23}
\times(-\half \gym^2) \Tr [T^a,T^d][T^b,T^c]
\lrbrk{\frac{X_{1234}}{I_{14}I_{23}}+F_{14,23}}.
\end{myeqnarray}
The scalar interaction is contained in $X$ and the gluon exchange in $F$.

\newpage
\subsection{Insertion into diagrams}
\label{ssec:levelzeroinsert}

The radiative corrections are obtained by decorating the
free theory diagrams with $\order{\gym^2}$ corrections.
Decoration means for every line insert the self-energy
and for every pair of lines insert a gluon exchange.
The scalar vertex and gluon exchange in 
\eqref{eq:PP.4ptG2} have the same algebraic structure 
and we refer to them collectively as gluon exchange.
Note that, when the radiative corrections 
are treated in this way, the genus of a diagram might be changed
due to a gluon line crossing a scalar.

\begin{figure}\centering
\[
\mathord{\parbox[c]{2.02cm}{\centering\includegraphics{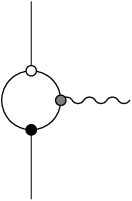}}}-
\mathord{\parbox[c]{2.02cm}{\centering\includegraphics{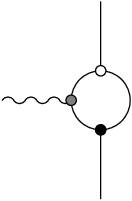}}},\qquad
\mathord{\parbox[c]{1.52cm}{\centering\includegraphics{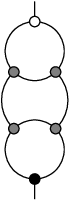}}}-
\mathord{\parbox[c]{1.52cm}{\centering\includegraphics{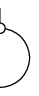}}}
\]
\caption{Algebraic structure of gluon vertex and scalar self-energy.}
\label{fig:PP.VertexZZG}
\end{figure}
Two scalars couple to a gluon by an
effective vertex which is proportional to $\Tr [Z,\bar Z]G=\Tr Z\bar Z G-\Tr \bar Z Z G$
(\figref{fig:PP.VertexZZG}).
This means a gluon can couple to the left or right
hand side of a scalar line and both possibilities are 
distinct and differ in sign.
\begin{figure}\centering
$\mathord{\parbox[c]{2.8cm}{\centering\includegraphics{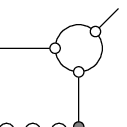}}}
-\mathord{\parbox[c]{2.8cm}{\centering\includegraphics{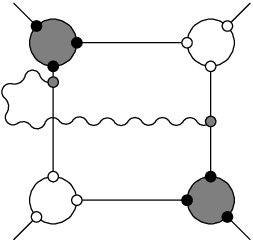}}}
-\mathord{\parbox[c]{2.8cm}{\centering\includegraphics{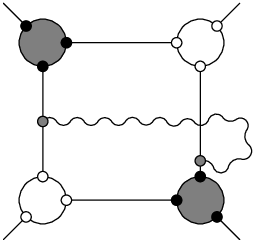}}}
+\mathord{\parbox[c]{2.8cm}{\centering\includegraphics{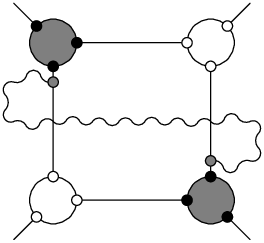}}}$
\\
$\mathord{\parbox[c]{2.8cm}{\centering $N$}}
\mathbin{\phantom{-}}\mathord{\parbox[c]{2.8cm}{\centering $1/N$}}
\mathbin{\phantom{-}}\mathord{\parbox[c]{2.8cm}{\centering $1/N$}}
\mathbin{\phantom{+}}\mathord{\parbox[c]{2.8cm}{\centering $1/N$}}$

\caption{Gluon exchange between two scalar lines.}
\label{fig:PP.GluonEx}
\end{figure}
If a gluon line is inserted between two edges of a
face of the diagram,
there is one way to insert the gluon without
crossing scalar lines and three ways 
where the gluon has to cross the scalar lines,
see \figref{fig:PP.GluonEx}. 
The planar insertion adds a face to the diagram
and thus has a factor of $N$,
the non-planar insertions 
require an additional handle and have a factor of $1/N$. 
As two of four insertions have a positive sign and two have a
negative sign, the sum of the four possibilities is $\pm(N-1/N)$. 
If the gluon line is inserted between lines which
are not edges of a common face, however, all
four gluon insertions require an additional handle and cancel.
Thus, gluons can only be exchanged between
the edges of a face, and we will only draw the planar 
insertion to represent the sum of all four.
The effective vertex for the self-energy 
is $N\Tr Z\bar Z-\Tr Z\Tr \bar Z$, see \figref{fig:PP.VertexZZG}
The first part of this vertex does not change
the graph, the second one breaks a line and
joins two faces. The sum of both 
has the combinatorial factor $N-1/N$.

To be more precise we now state the 
correction with space-time dependence and prefactors.
Assume we have a free theory diagram with value $A$. 
We insert a gluon exchange \eqref{eq:PP.4ptG2}
between the edges $x_1\to x_2$ and $x_3\to x_4$
of a face of the diagram.
The correction due to this gluon exchange diagram is
\begin{equation}\label{eq:PP.GluEx}
A'=\pm\frac{\gym^2(N^2-1)A}{2N}\lrbrk{\frac{X_{1234}}{I_{12}I_{34}}+F_{12,34}}.
\end{equation}
The sign is negative if the two lines have the same direction
on the boundary of the face and positive otherwise.
The self-energy \eqref{eq:PP.PropG2}
on the line $x_1\to x_2$ is actually
the same as a gluon exchange between the line $x_1\to x_2$ and itself,
i.e. \eqref{eq:PP.GluEx} with $x_3=x_1$ and $x_4=x_2$.
This can be seen by taking the limit
$x_3\to x_1$ and $x_4\to x_2$ of $F_{12,34}$ 
in \eqref{eq:PP.GluExSimp}.
In that limit some terms are cancelled by the fact
that $1/I_{12}$ vanishes quadratically at $x_1\to x_2$ 
while $X_{1234}$ and $Y_{123}$ only have logarithmic divergences. 

\paragraph{Corrections to a face.}

\begin{figure}\centering
$\frac{1}{2}\mathord{\parbox[c]{3.02cm}{\centering\includegraphics{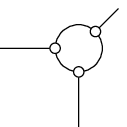}}}
-\mathord{\parbox[c]{3.02cm}{\centering\includegraphics{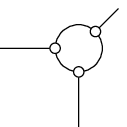}}}
+\mathord{\parbox[c]{3.02cm}{\centering\includegraphics{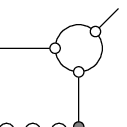}}}$
\\
$\mathord{\phantom{\frac{1}{2}}}\mathord{\parbox[c]{3.02cm}{\centering 4 perm}}
\mathbin{\phantom{-}}\mathord{\parbox[c]{3.02cm}{\centering 4 perm}}
\mathbin{\phantom{+}}\mathord{\parbox[c]{3.02cm}{\centering 2 perm}}$

\caption{Radiative corrections to a quadrangle.}
\label{fig:PP.FaceCorr}
\end{figure}

We now sum up all radiative corrections
within a face of a free diagram $A$,
see \figref{fig:PP.FaceCorr}.
The face is a $2n$-gon with vertices $x_0,x_1,\ldots,x_{2n}$, $x_0=x_{2n}$,
of alternating conjugation type.
The total radiative correction is
\begin{equation}
A'=-\frac{\gym^2(N^2-1)A}{4N}\sum_{k=1}^{2n}\sum_{l=1}^{2n} 
\lrbrk{F_{(k-1)k,(l-1)l}+(-1)^{k+l}\frac{X_{(k-1)k(l-1)l}}{I_{(k-1)k}I_{(l-1)l}}}.
\end{equation}
Note that the alternating sign of \eqref{eq:PP.GluEx} is
compensated by the anti-symmetry of $F_{12,34}$ 
in $x_1,x_2$ and $x_3,x_4$.
The gluon exchanges between different sides appear twice within the double sum,
which is compensated by a factor of $\half$ the prefactor.
The self-energies appear once in the double sum, but 
on two different faces, therefore the factor of $\half$ is correct here as well.
The contribution of the $G$s \eqref{eq:PP.GluExSimp} cancels
in the sum because the sum telescopes and is cyclic.
We can therefore write the correction as
\begin{myeqnarray}\label{eq:PP.FaceDeco}
A'\eq - \quarter \gym^2N(1-N^{-2})A\sum_{k=1}^{2n}\sum_{l=1}^{2n}
\\&&
\qquad\lrbrk{\frac{X_{(k-1)k(l-1)l}}{I_{(k-1)(l-1)}I_{kl}}-\frac{X_{(k-1)k(l-1)l}}{I_{(k-1)l}I_{k(l-1)}}+(-1)^{k+l}\frac{X_{(k-1)k(l-1)l}}{I_{(k-1)k}I_{(l-1)l}}}.
\nn
\end{myeqnarray}
The result is manifestly conformally invariant and finite.
For a bigon, $n=1$, the sum vanishes and for a quadrangle, $n=2$, the 
expression simplifies to 
\begin{equation}\label{eq:PP.Face4Deco}
A'=-(1-N^{-2})\gym^2N\,\frac{X_{1234}}{I_{13}I_{24}}\,A
=-(1-N^{-2})\,\frac{\gym^2N\Phi(r,s)}{(4\pi)^2}\,A.
\end{equation}
%

\paragraph{Special cases.}

The above discussion was not completely honest for two cases. 

Gluon corrections contribute only 
if they are between two edges of a common face. 
Sometimes, it may happen that a they are
also edges of another face. 
In that case, two of the four insertions of gluons
between the lines are planar and two are non-planar. 
They add up to a factor of $2(N-1/N)$ allowing for 
one normal gluon line in each face. 

There are diagrams  
with the following two equivalent properties.
There is a line that,
when removed, reduces the genus of the diagram.
The left and the right hand side of a (this) line can be
connected by a gluon without crossing the other lines.
For such a diagram two things happen.
In the above construction this line was
considered as two distinct sides of a face. 
The double-counting of gluons connecting this line
to some other line is correct, because the gluon couples
to both sides of the line. However, a gluon connecting the
right and left side of this line
should not have been taken into account.
Furthermore, the self-energy was claimed to have
a group factor of $N-1/N$. In this special case this is not so.
The broken line part 
$-\Tr Z\Tr \bar Z$ of the effective vertex 
does not contribute $-1/N$ here, but rather
$-N$, because one handle of the surface can be removed.
Thus, the self-energy should not be considered either
and the two erroneous contributions cancel.

In these two special cases the result is obtained
in the same way as described above.

\paragraph{Example: Extremal correlators.}

We consider the case of all operators of one kind at the 
same space-time point. This is called an \emph{extremal} correlator
\cite{D'Hoker:1999ea}.
Due to non-renormalization of extremal correlators
they must not receive radiative corrections.
It is easily seen that 
the summands in \eqref{eq:PP.FaceDeco}
vanish for coinciding points of the same color
confirming non-renormalization at one-loop order.

\newpage
\sect{Matrix model results on four-point functions}
\label{app:4ptmm}

Extremal correlators of chiral primaries of the type $\mbox{Tr}Z^J$
can be calculated for instance by using the result of
Ginibre~\cite{Ginibre},
as used already in \cite{Kristjansen:2002bb}.
For the four-point function the result reads (with $0< J_1,J_2,J_3,J_1+J_2+J_3<N$)
\begin{myeqnarray}
\lefteqn{
\langle {\rm Tr} Z^{J_1} {\rm Tr} Z^{J_2} {\rm Tr} Z^{J_3}{\rm Tr}\bar{Z}^{J_1+J_2+J_3}\rangle}
\nle
\frac{1}{J+1}
\left\{
\frac{\Gamma(N+J_1+J_2+J_3+1)}{\Gamma(N)}-
\frac{\Gamma(N+J_2+J_3+1)}{\Gamma(N-J_1)}
\right.
\nl
-\frac{\Gamma(N+J_1+J_3+1)}{\Gamma(N-J_2)}
-\frac{\Gamma(N+J_1+J_2+1)}{\Gamma(N-J_3)}
+\frac{\Gamma(N+J_1+1)}{\Gamma(N-J_2-J_3)}
\nl
\left.
+\frac{\Gamma(N+J_2+1)}{\Gamma(N-J_1-J_3)}
+\frac{\Gamma(N+J_3+1)}{\Gamma(N-J_1-J_2)}
-\frac{\Gamma(N+1)}{\Gamma(N-J_1-J_2-J_3)}\right\}.\qquad
\label{simple4pt}
\end{myeqnarray}
Based on the cases $n=2$~\cite{Kristjansen:2002bb,Constable:2002hw}, 
$n=3$~\cite{Kristjansen:2002bb} and $n=4$ it is natural to conjecture%
\footnote{
This was independently conjectured by K.Okuyama 
\cite{Okuyama:2002zn}. A proof should be straightforward using the 
Ginibre or character techniques, as in \cite{Kristjansen:2002bb}. }
that the general $n$-point function of this type takes the form 
(with $J\equiv\sum_{i=1}^n J_i$)
\begin{myeqnarray}
\lefteqn{
\langle {\rm Tr} Z^{J_1} {\rm Tr} Z^{J_2}\ldots 
{\rm Tr} Z^{J_n} {\rm Tr}\bar{Z}^{J}\rangle}
\nle
\frac{1}{J+1}
\left\{
\frac{\Gamma(N+J+1)}{\Gamma(N)}-
\sum_{i=1}^n \frac{\Gamma(N+J-J_i+1)}{\Gamma(N-J_i)}
\right.
\nl
+\sum_{1\leq i_1<i_2\leq n} \frac{\Gamma(N+J-J_{i_1}-J_{i_2}+1)}
{\Gamma(N-J_{i_1}-J_{i_2})}-\cdots
\left.
+(-)^n\frac{\Gamma(N+1)}{\Gamma(N-J)}\right\}.\qquad
\label{conjecture}
\end{myeqnarray}
Taking the double scaling limit of the expression~\eqref{simple4pt} we obtain
\begin{eqnarray}
\lefteqn{
\langle {\rm Tr} Z^{J_1} {\rm Tr} Z^{J_2} {\rm Tr} Z^{J_3}{\rm Tr}\bar{Z}^{J_1+J_2+J_3}\rangle
 }\label{dsl4pt} \nn\\
&&
\longrightarrow 4 N^{J}J\cdot
\frac{\sinh\left(\frac{J_1 J}{2N}\right)
\sinh\left(\frac{J_2 J}{2N}\right)
\sinh\left(\frac{J_3 J}{2N}\right)}
{\frac{J^2}{2N}}.
\end{eqnarray}
The double scaling limit of the conjectured formula~\eqref{conjecture} 
reads
\begin{equation}
\langle {\rm Tr} Z^{J_1} {\rm Tr} Z^{J_2}\ldots 
{\rm Tr} Z^{J_n} {\rm Tr}\bar{Z}^{J}\rangle
\longrightarrow 2^{n-1}(N^J J)\,
\frac{
\prod_{i=1}^n\sinh\left(\frac{J_i J}{2N}\right)}
{\frac{J^2}{2N}}.
\end{equation}

The non-extremal four-point function of chiral primaries of the type 
${\rm Tr} Z^J$ is harder to obtain entirely by matrix model calculations. 
However, for special configurations of space time points this four-point
function can be expressed in terms of a matrix model correlator which 
can be evaluated exactly, namely (with $0< {J_1},{J_2},{K_1},K_2={J_1}+{J_2}-{K_1}<N$)
\begin{eqnarray}
\lefteqn{\hspace{-1.5cm} \langle {\rm Tr} Z^{{J_1}} {\rm Tr} Z^{{J_2}} 
{\rm Tr} \bar{Z}^{{K_1}} {\rm Tr}\bar{Z}^{K_2}\rangle\indup{conn}= } \nonumber \\
& &\frac{1}{{J_1}+{J_2}+1} \left\{ \frac{(N+{J_1}+{J_2})!}{(N-1)!} \right. \nonumber \\
&&
+\frac{(N+\mathrm{Min}[{J_1},{J_2},{K_1},K_2])!}{(N-\mathrm{Max}[{J_1},{J_2},{K_1},K_2]-1)!} \nonumber \\
&&-\frac{(N+\mathrm{Max}[{J_1},{J_2},{K_1},K_2])!}{(N-\mathrm{Min}[{J_1},{J_2},{K_1},K_2]-1)!}
\nonumber \\
&& \left.-\frac{N!}{(N-{J_1}-{J_2}-1)!} \right\} \nonumber \\
&&-\sum_{\mathrm{Max}[0,{K_1}-{J_1}]}^{\mathrm{Min}[{J_2},{K_1}]-1}\frac{(N+m)!}{(N+m-{J_2})!}
\frac{(N+m+{J_1}-{K_1})!}{(N+m-{K_1})!}\nonumber \\
&&-\sum_{\mathrm{Max}[0,{K_1}-{J_2}]}^{\mathrm{Min}[{J_1},{K_1}]-1}\frac{(N+m)!}{(N+m-{J_1})!}
\frac{(N+m+{J_2}-{K_1})!}{(N+m-{K_1})!}.
\label{simplified}
\end{eqnarray}
{}From this expression it is 
straightforward to derive the genus expansion. Assuming ${J_1}\leq {J_2},{K_1}<K_2$ 
one finds
\begin{eqnarray}
\lefteqn{\hspace{-1.5cm} \langle {\rm Tr} Z^{{J_1}} {\rm Tr} Z^{{J_2}} 
{\rm Tr} \bar{Z}^{{K_1}} {\rm Tr}\bar{Z}^{K_2}\rangle\indup{conn}= }\nonumber \\ 
&&N^{{J_1}+{J_2}} \sum_{p=0}^{\infty}\frac{1}{N^p} \left(
\alpha_{p+1}-2 \sum_{i=0}^p \sum_{m=0}^{{J_1}-1} \beta_i(m)\, \gamma_{p-i}(m)
\right)
\label{genusexp}
\end{eqnarray}
where
\begin{myeqnarray}
\alpha_p\eq\frac{1}{{J_1}+{J_2}+1}\, \sum_{0\leq i_1< \ldots< i_p\leq {J_1}+{J_2}}
\left[ (1-(-1)^p)\prod_{q=1}^p i_q \right. 
\nl
\hspace{1.2cm}\left.+\prod_{q=1}^p (i_q-{J_2})
+\prod_{q=1}^p (i_q-{J_1})\right], \hspace{0.5cm} p\geq1
\end{myeqnarray}
and
\begin{myeqnarray}
\beta_0(m)\eq\gamma_0(m)=1,
\nn \\
\beta_p(m)\eq\sum_{0\leq i_1< \ldots< i_p\leq {J_1}-1}
\left[\prod_{q=1}^p(m-i_q)\right],\hspace{0.5cm} p\geq 1, \nonumber \\
\gamma_p(m)\eq\frac{1}{2}\,\sum_{0\leq i_1< \ldots< i_p\leq {J_2}-1}
\left[\prod_{q=1}^p(m+{K_1}-{J_1}-i_q)\right. 
\nl
\qquad\left.+\prod_{q=1}^p(m+{J_2}-{K_1}-i_q)\right],
\hspace{0.5cm}p\geq 1.
\end{myeqnarray}
{}From here we can generate explicit expressions for 
(in principle) any term in the genus expansion. 
For lower genera this results in
\begin{eqnarray}
\lefteqn{ \langle {\rm Tr} Z^{{J_1}} {\rm Tr} Z^{{J_2}} 
{\rm Tr} \bar{Z}^{{K_1}} {\rm Tr}\bar{Z}^{K_2}\rangle\indup{conn}= } \\ 
&& N^{{J_1}+{J_2}-2}{J_1}{J_2}{K_1}K_2({J_2}-1)\left\{1+
\frac{1}{24 N^2}
\left[{J_1}^4+2{J_1}^3({J_2}-3)\right. \right.\nonumber \\
&&+{J_1}^2(11+2({J_2}-5){J_2}) 
+2{J_1}({J_2}-3)({J_2}^2+2{K_1}-{J_2}(2+{K_1})-1)
\nonumber \\
&&\left.\left.
+({J_2}-3)({J_2}-2)({J_2}({J_2}-1)-2-2{J_2}{K_1}+2{K_1}^2)
\right]
+{\cal O}\left(\frac{1}{N^4}\right)\right\}.\nn
\end{eqnarray}
Likewise, we can easily take our double scaling limit and
we get
\begin{eqnarray}
\lefteqn{
\langle {\rm Tr} Z^{J_1} {\rm Tr} Z^{J_2} {\rm Tr} \bar{Z}^{K_1}{\rm Tr}\bar{Z}^{K_2}
\rangle\indup{conn} }\\
&& \longrightarrow 4 N^{{J_1}+{J_2}}({J_1}+{J_2})\frac{\sinh\left(\frac{({J_1}+{J_2}){J_1}}{2N}\right)
\sinh\left(\frac{{J_2} {K_1}}{2N}\right)\sinh\left(\frac{{J_2} K_2}{2N}\right)}
{\frac{({J_1}+{J_2})^2}{2N}}. \nonumber
\end{eqnarray}

The above results can alternatively be derived by character expansion 
techniques \cite{Kostov:1997bs,Kostov:1998bn}, as in 
\cite{Kristjansen:2002bb}. It is interesting to note that for the
non-extremal correlator eq.\eqref{simple4pt} one needs to
consider double-hook Young diagrams, as opposed to the single-hook
diagrams sufficient for extremal, free correlators.

\sect{Effective vertices for one-loop two-point functions}
\label{app:effver}

We would like to turn the calculation of U($N$) $\superN=4$ SYM 
two-point functions of BMN operators 
into a matrix model problem. 
For that we denote the scalar fields at the space-time points $0$
and $x$ by $\phi^-_i:=\phi_i(0)$ and $\phi^+_i:=\phi_i(x)$, respectively. 
The free SYM correlators are 
\begin{equation}
\bigvev{\phi^{-(a)}_i \phi^{+(b)}_j}=\frac{\gym^2\,\delta_{ij}\delta^{ab}}{8\pi^2 x^2},
\qquad
\bigvev{\phi^{-(a)}_i \phi^{-(b)}_j}=\bigvev{\phi^{+(a)}_i \phi^{+(b)}_j}=0,
\end{equation}
the latter two giving rise to tadpoles which are zero in dimensional 
regularization. With these rules one can compute 
free correlators $\vev{O_1^- O_2^{+}}$ of the operators
$O_1^-:=O_1(0)$ and $O_2^{+}:=O_2(x)$ consisting of SYM scalars.
As each scalar field $\phi^-_i$ must be 
contracted with a field $\phi^+_j$, the number of factors
of $\gym^2/8\pi^2 x^2$ is known and will be dropped for the sake 
of simplicity. What remains is a canonically normalized 
Gaussian matrix model with the U($N$) contraction rules
\begin{myeqnarray}
\bigbrk{\Tr (\phi^-_i A) \Tr (\phi^+_j B)}_{\phi^-_i\,\leftrightarrow\, \phi^+_j}
\eq\delta_{ij}\,\Tr AB,
\nn\\
\bigbrk{\Tr (\phi^-_i A \phi^+_j B)}_{\phi^-_i\,\leftrightarrow\, \phi^+_j}\eq
\delta_{ij}\,\Tr A \Tr B,
\end{myeqnarray}
and all other contractions zero.

SYM interactions can be included in the matrix model 
by adding effective vertices, which represent the combinatorial
structure of the SYM interactions. The space-time integrals 
of the interactions, however, need to be computed by hand and appear 
in the coupling constant of the effective vertex.

At one-loop, there are three kinds of interactions of interest,
scalar self-energies, gluon-exchanges and scalar vertices.
The scalar interaction term of $\superN=4$ SYM 
(\emph{cf} the action in eq.\eqref{symaction})
\begin{equation}
U=-\frac{1}{2\gym^2}\Tr [\phi_i,\phi_j][\phi_i,\phi_j].
\end{equation}
couples to either of the operators
at $0$ or $x$. We therefore set
$\phi_i=\phi^+_i+\phi^-_i$
and isolate the part with two $\phi^-$ and two $\phi^+$
\begin{equation}
U=
-\frac{1}{\gym^2}
\lrbrk{\Tr [\phi^+_i,\phi^-_j][\phi^+_i,\phi^-_j]
+\Tr [\phi^+_i,\phi^-_j][\phi^-_i,\phi^+_j]
+\Tr [\phi^+_i,\phi^+_j][\phi^-_i,\phi^-_j]}
\end{equation}
as we are interested only in those interactions that 
preserve the number of fields.
Using a Jacobi identity the potential can be split up into three parts
\begin{myeqnarray}
U\eq \frac{2}{\gym^{2}} (V_D+V_F+V_K),
\nn\\
V_D\eq\half\Tr [\phi^+_i,\phi^-_i][\phi^+_j,\phi^-_j],
\nn\\
V_F\eq-\Tr [\phi^+_i,\phi^+_j][\phi^-_i,\phi^-_j],
\nn\\
V_K\eq-\half\Tr [\phi^+_i,\phi^-_j][\phi^+_i,\phi^-_j].
\end{myeqnarray}
The coupling constant of the corresponding effective vertices is given
by
\begin{equation}
-\frac{\gym^4 x^4}{(8\pi^2)^2}\int \frac{d^4 z}{(x-z)^4\,z^4} 
=\frac{\gym^4 L}{32\pi^2}.
\end{equation}
where it is understood that we have scaled away
the tree-level $x$-dependence of correlators.
In regularization by dimensional reduction we have
\begin{equation}
L=\log x^{-2}-\lrbrk{\frac{1}{\epsilon}+\gamma+\log \pi+2}.
\label{L}
\end{equation}
Clearly $\log \Lambda^2=\lrbrk{\frac{1}{\epsilon}+\gamma+\log \pi+2}$
is a (divergent) constant setting the scale, see discussion 
around eq.\eqref{extract}.
The effective vertex for the scalar interaction is
\begin{equation}
\frac{\gym^2 L}{16\pi^2}\bigbrk{\normord{V_D}+\normord{V_F}+\normord{V_K}}.
\end{equation}
The vertices for scalar self-energy
\begin{equation}
\frac{\gym^2 (L+1)}{8\pi^2}
 \bigbrk{N\normord{\Tr (\phi^-_i\phi^+_i)}-\normord{\Tr \phi^-_i \,\Tr \phi^+_i}}
\end{equation}
and gluon-exchange
\begin{equation}
\frac{\gym^2(L+2)}{32\pi^2} \bigbrk{\normord{\Tr [\phi^+_i,\phi^-_i][\phi^+_j,\phi^-_j]}},
\end{equation}
are obtained in a similar fashion.

\paragraph{Cancellation of D-terms.}

The term $V_D$ in the scalar interaction
cancels against the gluon-exchanges and
the scalar self-energies.
This fact was already efficiently used in \cite{Constable:2002hw}.   
The proof goes as follows. 
The sum of these terms can be written 
without normal-orderings in the following way
\begin{equation} \label{D-terms}
\frac{\gym^2 (L+1)}{8\pi^2}
\lrbrk{\half \Tr[\phi^+_i,\phi^-_i][\phi^+_j,\phi^-_j]
-N\Tr \phi^+_j \phi^-_j+\Tr \phi^+_j\,\Tr\phi^-_j
}.
\end{equation}
It is easy to see that $\phi^+_i$ in the above vertex cannot be contracted with an arbitrary trace
of scalars $\phi^-_{i_k}$
\begin{myeqnarray}
\lefteqn{\lrbrk{\Tr ([\phi^+_i,\phi^-_i][\phi^+_j,\phi^-_j])\Tr (\phi^-_{i_1} \phi^-_{i_2}\cdots \phi^-_{i_n})}
_{\phi^+_i\leftrightarrow \phi^-_{i_k}}}
\nle
\Tr \bigbrk{\bigcomm{\phi^-_{i_1}}{[\phi^+_j,\phi^-_j]}\phi^-_{i_2}\cdots \phi^-_{i_n}}
+\Tr \bigbrk{\phi^-_{i_1} \bigcomm{\phi^-_{i_2}}{[\phi^+_j,\phi^-_j]}\cdots \phi^-_{i_n}}
+\ldots
\nle
-\Tr \bigbrk{[\phi^+_j,\phi^-_j] \phi^-_{i_1}\phi^-_{i_2}\cdots \phi^-_{i_n}}
+\Tr \bigbrk{\phi^-_{i_1} \phi^-_{i_2}\cdots \phi^-_{i_n}[\phi^+_j,\phi^-_j]}=0.
\end{myeqnarray}
due to a telescoping sum and cyclicity of the trace. 
Furthermore, terms resulting from contracting the $\phi_i^+$ with one of
the $\phi^-$ inside the same vertex cancel against the remaining terms
in~\eqref{D-terms}. Thus the combination~\eqref{D-terms} does not give
any contribution to  two-point correlators of scalar fields.

\paragraph{F and K terms.}

The F-terms couple to anti-symmetric pairs of scalars and
the K-terms couple to traces. Hence symmetric traceless operators
do not couple to F and K terms and do not receive radiative corrections
at $\order{\gym^2}$.
We combine $\phi^5$ and $\phi^6$ to the complex field 
$Z=(\phi_5+i\phi_6)/\sqrt{2}$ and assume for $\phi_i$ that $i\leq 4$
\begin{myeqnarray}
V_F\eq
-\Tr [\phi^+_i,\phi^+_j][\phi^-_i,\phi^-_j]
-2\Tr[Z^+,\Zb^+][\Zb^-,Z^-]
\nl\qquad\mathord{}
-2\Tr[Z^+,\phi^+_i][\Zb^-,\phi^-_i]
-2\Tr[\Zb^+,\phi^+_i][Z^-,\phi^-_i]
\nn\\
V_K\eq-\half \Tr [\phi^+_i,\phi^-_j][\phi^+_i,\phi^-_j]
\nn\\&&\qquad\mathord{}
-\Tr [Z^+,\phi^-_i][\Zb^+,\phi^-_i]
-\Tr [\phi^+_i,Z^-][\phi^+_i,\Zb^-]
\nl\qquad\mathord{}
-\Tr [Z^+,Z^-][\Zb^+,\Zb^-]
-\Tr [Z^+,\Zb^-][\Zb^+,Z^-].
\end{myeqnarray}
The one-loop 
expectation value of a two-point correlator is obtained by gluing in
the effective scalar interaction
\begin{equation}
\delta_{\gym^2}\vev{O_1(0) O_2(x)}=\frac{\gym^2 L}{16\pi^2}\,\bigvev{(\normord{V_F}+\normord{V_K}) \, O_1^- O_2^+}.
\label{master}
\end{equation}
For calculations of two-point functions
we may set $L=\log x^{-2}$, as the remaining 
divergent and finite parts are always the same and can be absorbed into 
redefinitions of the fields~\cite{Gross:2002su}.

\sect{Diagrammatic computation of correlators}
\label{app:diagram}

\paragraph{Free Correlators.}

In this section we present a diagrammatic way 
to determine correlators of BMN operators and
the diagrams involved in the evaluation 
of eqs.\eqref{startingpoint},\eqref{doublecorrs},\eqref{double-single}.
It is similar to methods applied in \cite{Constable:2002hw,Chu:2002pd}.

\begin{figure}
\centering
\includegraphics{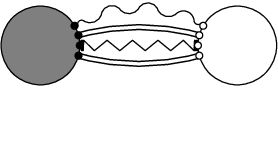}
\qquad
\includegraphics{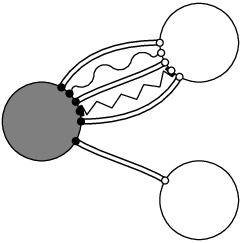}
\includegraphics{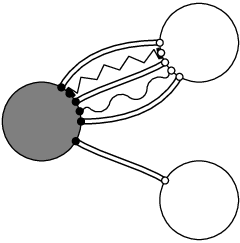}\qquad
\includegraphics{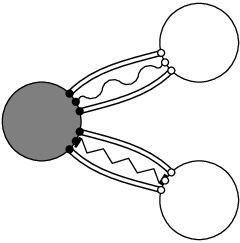}\vspace{0.5cm}

\includegraphics{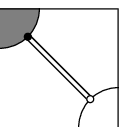}
\includegraphics{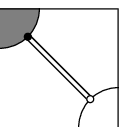}
\includegraphics{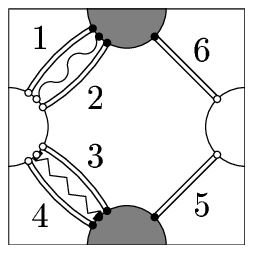}
\includegraphics{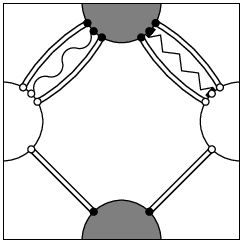}
\includegraphics{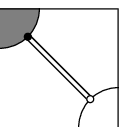}
\caption{Free correlator of BMN operators. 
The first diagram on the top line is the sphere of
$\vev{\bar\cO^J_{12,n}\cO^{J}_{12,m}}$,
the middle two are 
$\vev{\bar\cO^J_{12,n}\cT^{J,r}_{12,m}}$
and the last one is 
$\vev{\bar\cO^J_{12,n}\cT^{J,r}_{12}}$.
The diagrams on the bottom line 
are the torus of
$\vev{\bar\cO^J_{12,n}\cO^{J}_{12,m}}$
on a periodic square.}
\label{fig:PP.FreeDiag}
\end{figure}

We represent the traces of an operator by circles, 
the empty circles are composed mostly of $Z$s,
the shaded ones mostly of $\bar Z$s.
In the free theory the fields in the traces are connected by 
lines. Plain lines connect $Z$s to $\bar Z$s, 
wiggly or zigzag lines connect two impurities $\phi_1$ or $\phi_2$,
respectively. 
The majority of lines are plain lines and 
on surfaces of low genus most of these lines run parallel to another. 
To make the diagrams more concise we draw a bunch of parallel plain lines
(see \figref{fig:PP.22Sphere})
as one double line. The number of constituent lines will be 
denoted by $a_k$ where $k$ is the label of the double line
determined as follows:
We consider the leftmost circle and label the double lines 
starting at the top in clockwise order from $1$ to the number of
double lines.
The value of a diagram is the sum over all possible 
sizes $a_k$ of double lines weighted with the phase factors of
the BMN operators. 
For each circle the total number of lines must 
equal the number of fields on the trace, this
is achieved by inserting a Kronecker delta 
into the sum. Furthermore, for an operator with 
two impurities, charge $J$ and mode number $n$
we insert the phase $\exp(2\pi i b n/J)$, where $b$ is
the distance from the wiggly to the zigzag line in 
clockwise direction. Finally, we have to 
multiply by the normalization factors from the
definition of the operators eqs.\eqref{single},\eqref{double}.
For example the third torus diagram in \figref{fig:PP.FreeDiag}
has the value
\begin{equation}
\frac{1}{J}\sum_{a_1,\ldots,a_6=1}^{J}
\delta_{a_1+\ldots+a_6,J}\,
\exp \frac{2\pi i n (a_2+a_3)}{J}\,
\exp \frac{2\pi i m (a_1+a_4)}{J}.
\end{equation}
This expression has the correct leading $J$ behavior,
we can therefore transform the sum into an integral
\begin{equation}
\frac{1}{J}\int_0^J d^6 a\,
\delta(a_1+\ldots+a_6-J)\,
\exp \frac{2\pi i n (a_2+a_3)}{J}\,
\exp \frac{2\pi i m (a_1+a_4)}{J},
\end{equation}
as we are not interested in $\order{1/J}$ corrections in this work.
It turns out that for all relevant diagrams 
the corresponding sums can be approximated by integrals in the BMN limit.
All diagrams that contribute to the correlators in 
eqs.\eqref{startingpoint},\eqref{doublecorrs},\eqref{double-single} 
are shown in \figref{fig:PP.FreeDiag}.

\paragraph{Radiative Corrections.}

\begin{figure}
\centering
\includegraphics{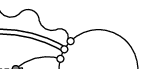}\quad
\includegraphics{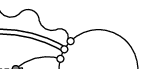}\quad
\includegraphics{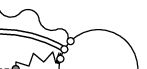}\quad
\includegraphics{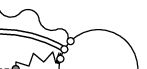}\\
\makebox[3.05cm][c]{$+$}
\makebox[3.05cm][c]{$-$}
\makebox[3.05cm][c]{$-$}
\makebox[3.05cm][c]{$+$}
\caption{F-term interactions in 
$\vev{\bar\cO^J_{12,n}\cO^{J}_{12,m}}$
on the sphere.
Another four diagrams have the two impurities interchanged.}
\label{fig:PP.2ptSphereF}
\end{figure}

The radiative corrections to the correlators are 
obtained by inserting the effective vertex eq.\eqref{master}
into the matrix model correlator. 
The F-terms couple anti-symmetrically to an operator
while the K-terms couple only to SO(6) traces.
Only the singlet BMN operator has an SO(6) trace
and the extra contributions will be calculated later.
For the time being we would like to concentrate on the F-term.
The relevant part of the $F$-term effective vertex is 
$(-\gym^2 L/8\pi^2) \Tr [\bar Z^+,\phi_i^+][Z^-,\phi_i^-]$.
The trace can be separated into two traces linked by a line
$\Tr [\bar Z^+,\phi_i^+]T^a \,\Tr [Z^-,\phi_i^-]T^a $.
Graphically we thus represent the four-point scalar interactions
by two three-point interactions joined by a dashed line. 
We will suppress the factor $(-\gym^2L/8\pi^2)$ at intermediate stages
and put it back in the end.

First of all, we will consider the sphere of the correlator of two
BMN operators, 
$\vev{\bar\cO^J_{12,n}\cO^{J}_{12,m}}$, see \figref{fig:PP.2ptSphereF}.
In the first two diagrams the distance between the 
wiggly and zigzag line is $a_1$ and $a_1+1$, respectively. 
Furthermore, the two diagrams receive different signs from 
the effective vertex, a plus for the first, a minus for the second.
Therefore the sum of both diagrams receives the effective phase factor
\begin{equation}
\exp\frac{2\pi i a_1 n}{J}-\exp\frac{2\pi i (a_1+1) n}{J}
=
-2i\exp\frac{2\pi i (a_1+\half) n}{J}\,\sin \frac{\pi n}{J}
\end{equation}
from the left operator with mode number $n$.
In the BMN limit this simplifies to
\begin{equation}
-\frac{2\pi i n}{J}\exp\frac{2\pi i a_1 n}{J}.
\end{equation}
This can be generalised: When adjacent plain and zigzag lines 
connect a BMN operator with mode number $n$ and charge $J$ 
to an F-term vertex, the sum of the two possible
contributions is $-2\pi i n/J$ times the contribution 
where the zigzag line comes first in clockwise order. 
For the interactions between plain and wiggly lines 
the factor is $+2\pi i n/J$. 
The sum of the four depicted diagrams is thus
$(-2\pi i n/J)(-2\pi i m/J)$ times the 
free result. Together with the four diagrams
where the wiggly line interacts and the 
prefactor $(-\gym^2 L/8\pi^2)$ the 
total one-loop result is  $\lambda' L nm$ times the free
result.

\begin{figure}
\centering
\includegraphics{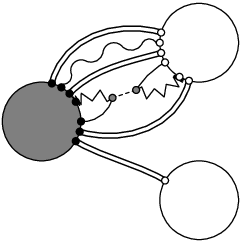}
\includegraphics{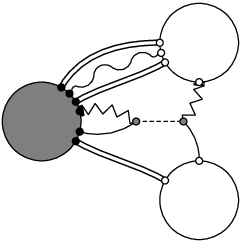}
\includegraphics{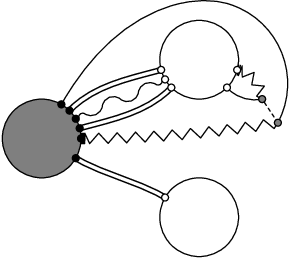}\qquad\qquad
\includegraphics{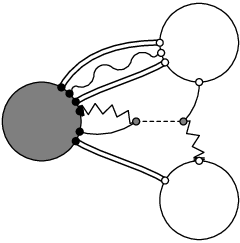}\vspace{0.5cm}

\includegraphics{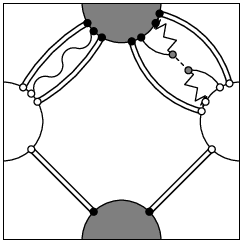}
\includegraphics{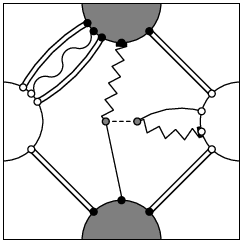}
\includegraphics{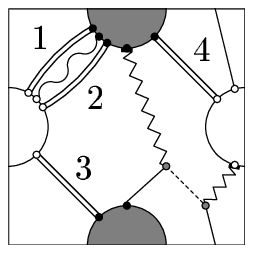}
\caption{F-term interactions in correlators of BMN operators. 
The first three belong to the correlator
$\vev{\bar\cO^J_{12,n}\cT^{J,r}_{12,m}}$,
the next one to
$\vev{\bar\cO^J_{12,n}\cT^{J,r}_{12}}$
and the last three to the torus of
$\vev{\bar\cO^J_{12,n}\cO^{J}_{12,m}}$
on a periodic square.}
\label{fig:PP.OneLoopF}
\end{figure}

\Figref{fig:PP.OneLoopF} contains the diagrams 
that contribute to the remaining correlators in
eqs.\eqref{startingpoint},\eqref{doublecorrs},\eqref{double-single}.
We have shown only representative diagrams, there are several 
other diagrams with different positions of the impurities 
and different orientations of inserted vertices.
The first depicted diagrams on both lines 
contribute $\lambda' L n m$ times the free result 
of the corresponding correlator for the same reason as before.
The other diagrams can also be shown to be proportional to 
the free result except the last one on the torus. 
As an example we will write down the phase factor from this 
particular diagram in the case $n=m$
\begin{equation}
-\exp\frac{2\pi i(a_2+a_3)n}{J}\,
\exp\frac{2\pi i(a_1+a_3+a_4)n}{J}
\end{equation}
where the two phases can be combined to $2\pi a_3 n/J$.
Adding the three diagrams where the zigzag line is interchanged with 
the plain lines on the interaction we get
\begin{equation}
-\exp\frac{2\pi i a_3 n}{J}
\lrabs{1-\exp\frac{2\pi i a_4 n}{J}}^2.
\end{equation}
Then we add the cases where the wiggly line sits 
on one of the other two edges formed by double lines
\begin{equation}
-\lrabs{1-\exp\frac{2\pi i (a_3+a_4) n}{J}}^2
-\exp\frac{-2\pi i a_1 n}{J}
\lrabs{1-\exp\frac{2\pi i a_4 n}{J}}^2.
\end{equation}
This we must multiply by $2$ for interchange of
impurities, the prefactor $(-\gym^2 L/8\pi^2)$ and
the normalization $1/J$.
After integration over the $a_k$ we get
\begin{equation}
\frac{\gtwo^2 \lambda' L}{8\pi^2}\lrbrk{\frac{2}{3}+\frac{5}{\pi^2 n^2}}=
\frac{\gtwo^2 \lambda' L}{8\pi^2}\mathcal{D}^1_{nn}.
\end{equation}

\newpage
\paragraph{Singlet Operators and K-term Interactions.}

\begin{figure}
\centering
\includegraphics{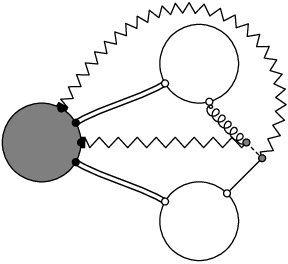}\qquad
\includegraphics{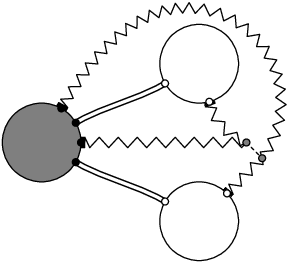}\qquad
\includegraphics{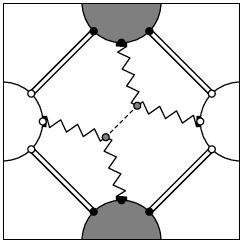}
\caption{K-term interactions of singlet BMN operators.
They contribute to 
$\vev{\bar\cO^J_{\singlet,n}\cT^{J,r}_{\singlet,m}}$,
$\vev{\bar\cO^J_{\singlet,n}\cT^{J,r}_{\singlet}}$
and the torus of
$\vev{\bar\cO^J_{\singlet,n}\cO^{J}_{\singlet,m}}$.}
\label{fig:PP.SingletK}
\end{figure}

For the singlet operator $\cO^J_{\singlet,n}$ there are additional contributions
at one-loop due to the K-term interaction which couples
to SO(6) traces. 
Assume the distance between the $\phi_k$ is $b$.
Then the associated phase factor is $4\cos 2\pi i n b/J$ explained 
as follows: The definition of the singlet operator involves 
a factor of $\half$. There is a contribution from 
each of the $4$ scalar flavors.
Furthermore, there is no distinction between the
two impurities and we must add up two 
conjugate complex phases giving a $2\cos$.
Our definition of the singlet operator, however,
involves also the piece $-\Tr \bar Z Z^{J+1}$ 
(a line between $\bar Z$ and $Z$ in the opposite 
direction is drawn as a curly line).
The strength of this contribution was adjusted to add
$-4$ to the above phase factor which gives the total effective phase factor
\begin{equation}
-4+4\cos \frac{2\pi i n b}{J}=-8\sin^2 \frac{\pi n b}{J}.
\end{equation}
This immediately shows that there is no interaction 
for nearby impurities, $b=0$, due to the K-term
and this reduces the number of contributing diagrams.

The only contributions to the correlators under consideration
are due to the diagrams in \figref{fig:PP.SingletK}.
Let us consider the first diagram. 
There is a phase factor of $-8\sin^2 \pi n b/J$ from the 
coupling to $\bar\cO^J_{\singlet,n}$.
The vertex couples to $\cT^{J,r}_{\singlet,m}$ 
only with $\bar Z$ and $Z$ which gives a factor of $-4$.
Then, there is a crossed diagram with a different 
orientation of the dashed line which gives the same contribution.
Finally this needs to be multiplied by the effective
vertex prefactor $(-\gym^2 L/32\pi^2)$ and the total 
prefactor is $-16 \lambda' L/8\pi^2$ as in eqs.\eqref{singcorrs}.
The prefactor for the second diagram is the same, except
that the effective vertex couples to 
the two $\phi_k$ of $\cT^{J,r}_{\singlet}$ instead of $Z$ and $\bar Z$ of 
$\cT^{J,r}_{\singlet,m}$ and this amounts to a relative sign.

\sect{Matrix elements}
\label{app:matel}

The matrix elements that appear in the computation 
of correlators.
$M^1_{mn}$ and $\mathcal{D}^1_{mn}$
were already calculated in \cite{Kristjansen:2002bb,Constable:2002hw}.
Assume $|m| \neq |n|$, $m\neq 0$, $n \neq 0$.
\begin{myeqnarray}
M^1_{nn}\eq\frac{1}{60}
-\frac{1}{24\pi^2n^2}
+\frac{7}{16\pi^4n^4}
\nn\\
M^1_{n,-n}\eq\frac{1}{48\pi^2n^2}+\frac{35}{128\pi^4n^4}
\nn\\
M^1_{mn}\eq\frac{1}{12\pi^2(n-m)^2}
-\frac{1}{8\pi^4(n-m)^4}
\nl
+\frac{1}{4\pi^4m^2n^2}
+\frac{1}{8\pi^4mn(n-m)^2}
\nn\\\nn\\
M^2_{nn}\eq-\frac{1}{40}+\frac{1}{12\pi^2 n^2}-\frac{49}{128\pi^4n^4}
\nn\\
M^2_{n,-n}\eq-\frac{1}{96\pi^2n^2}-\frac{35}{256\pi^4n^4}
\nn\\
M^2_{mn}\eq-\frac{1}{8\pi^2(m-n)^2}
+\frac{3}{16\pi^4(m-n)^4}
+\frac{1}{8\pi^4(m+n)^2(m-n)^2}
\nl
-\frac{3}{16\pi^4m^2 n^2}
-\frac{3}{16\pi^4mn(m-n)^2}
\nn\\\nn\\
\mathcal{D}^1_{nn}=\mathcal{D}^1_{n,-n}\eq\displaystyle\frac{2}{3}+\frac{5}{\pi^2n^2}
\nn\\
\mathcal{D}^1_{mn}\eq\frac{2}{3}+\frac{2}{\pi^2 m^2}+\frac{2}{\pi^2 n^2}
\nn\\\\
\mathcal{D}^2_{nn}=\mathcal{D}^2_{n,-n}\eq\displaystyle-\frac{1}{2}-\frac{45}{16\pi^2n^2}
\nn\\
\mathcal{D}^2_{mn}\eq-1-\frac{3}{2\pi^2 n^2}-\frac{3}{2\pi^2 m^2}
+\frac{3}{4\pi^2 (n-m)^2}+\frac{3}{4\pi^2 (n+m)^2}
\nn
\end{myeqnarray}

\newpage


\bibliography{Mixing}
\bibliographystyle{Mixing}


\end{document}